\documentclass[lettersize,journal]{IEEEtran}
\usepackage{amsmath,amsfonts}
\usepackage{algorithmic}
\usepackage{array}
\usepackage{textcomp}
\usepackage{stfloats}
\usepackage{url}
\usepackage{verbatim}
\usepackage{graphicx}
\usepackage{bm}
\usepackage{cite}
\usepackage{cases}
\usepackage{booktabs}
\usepackage{subcaption}
\usepackage{url}

\def\BibTeX{{\rm B\kern-.05em{\sc i\kern-.025em b}\kern-.08em
    T\kern-.1667em\lower.7ex\hbox{E}\kern-.125emX}}
\usepackage{balance}
\begin{document}
\title{BeamformNet: Deep Learning-Based Beamforming Method for DoA Estimation~via~Implicit~Spatial~Signal~Focusing~and~Noise~Suppression}
\author{Xuyao Deng, Yong Dou, Kele Xu\textsuperscript{*},~\IEEEmembership{Senior Member,~IEEE}
\thanks{All authors are with the College of Computer Science and Technology, National University of Defense Technology, Changsha 470000 China (e-mail: \{dengxuyao, yongdou, xukelele\}@nudt.edu.cn). *Corresponding author.}}

% \markboth{Journal of \LaTeX\ Class Files,~Vol.~18, No.~9, September~2020}%
% {How to Use the IEEEtran \LaTeX \ Templates}

\maketitle

\begin{abstract}
%Deep learning–based direction-of-arrival (DoA) estimation has gained increasing popularity. A popular family of DoA estimation algorithms is beamforming methods, which operate by constructing a spatial filter that is applied to array signals. However, these spatial filters obtained by traditional model-driven beamforming algorithms fail under demanding conditions such as coherent sources and a small number of snapshots. In order to obtain a robust spatial filter, this paper proposes BeamformNet—a novel deep learning framework grounded in beamforming principles. Based on the concept of optimal spatial filters, BeamformNet leverages neural networks to approximately obtain the optimal spatial filter via implicit spatial signal focusing and noise suppression, which is then applied to received signals for spatial focusing and noise suppression, thereby enabling accurate DoA estimation. Experimental results on both simulated and real-world speech acoustic source localization data demonstrate that BeamformNet achieves state-of-the-art DoA estimation performance and has better robustness.
Direction-of-arrival (DoA) estimation is a cornerstone of array signal processing, with beamforming methods representing a prominent family of techniques that construct spatial filters to enhance target signals while suppressing noise and interference. Traditional model-driven beamformers, such as Conventional Beamforming (CBF) and Minimum Variance Distortionless Response (MVDR), excel under ideal conditions but degrade in challenging scenarios involving coherent sources, limited snapshots, array mismatches, or low signal-to-noise ratios (SNRs). To address these limitations, we introduce BeamformNet, a novel deep learning framework that leverages recurrent neural networks (RNNs) and attention mechanisms to approximate an optimal spatial filter, grounded in beamforming principles and signal sparsity in the spatial domain.
BeamformNet implicitly performs spatial signal focusing and noise suppression by processing the array manifold and received signals through bidirectional RNNs for contextual feature extraction, followed by an attention module for adaptive fusion and a projection layer to generate the spatial filter. This filter is applied to the received signals to yield a sparse spatial energy spectrum, enabling accurate DoA estimation via peak search after thresholding. Theoretically, we prove the existence of the optimal spatial filter under assumptions of narrowband far-field signals and fewer sources than array elements, with neural networks approximating it via the universal approximation theorem. The framework formulates the optimal spatial filter approximation as a multi-label classification task, trained using asymmetric loss to handle class imbalance.
Extensive experiments on simulated datasets and real-world benchmarks demonstrate that BeamformNet achieves state-of-the-art performance, outperforming methods like DA-MUSIC, DeepSSE, SubspaceNet, IQResNet, LowSNRNet, and classical MUSIC across varying numbers of sources (K), snapshots (T), SNRs, angular separations, array elements (M), and mismatches. It exhibits superior generalization and robustness, particularly under coherent sources and limited snapshots. BeamformNet's parameter count scales modestly with M, ensuring efficiency. Code is open-sourced on GitHub to advance the community~\footnote{\url{https://github.com/colaudiolab/BeamformNet}}.
\end{abstract}


\section{Introduction}

\IEEEPARstart{D}{irection} of arrival (DoA) estimation, a fundamental task in array signal processing, aims to locate sources using signals received by an array of sensors~\cite{benesty2017fundamentals}. Classical model-driven methods achieve strong performance under ideal conditions~\cite{barabell1983improving,bienvenu1983optimality,schmidt1986multiple,roy2002esprit,capon2005high,malioutov2005sparse}. However, in practice, these methods suffer from performance degradation due to signal coherence, array mismatch, limited snapshots, and so on~\cite{krim1996two,tuncer2009classical}. Recently, deep learning has advanced DoA estimation owing to its data-driven nature~\cite{papageorgiou2021deep,lee2022deep,merkofer2022deep,gao2023gridless,merkofer2023music,shmuel2023deep,chen2024sdoa,shmuel2024subspacenet,xu2025deep,gast2025dcd}. Despite their black-box nature, high complexity, and limited interpretability, neural networks effectively handle non-ideal conditions where traditional model-driven methods struggle, demonstrating greater robustness in demanding scenarios. How to combine interpretability and robustness in complex scenarios has become a key point in the field of neural network-based DoA estimation.

Beamforming methods are a popular family of traditional model-driven DoA estimation algorithms that are limited by model assumptions and are difficult to adapt to complex scenarios, including Conventional BeamForming (CBF)~\cite{krim1996two}, Minimum Variance Distortionless Response (MVDR)~\cite{capon2005high} and so on. At its core, beamforming constructs a spatial filter applied to array signals to estimate DoA~\cite{van1988beamforming}. Different beamforming algorithms are essentially derived from distinct optimization objectives. For instance, MVDR beamformer aims to minimize the total output power while maintaining a unity gain in the direction of interest, thereby suppressing interference and noise~\cite{capon2005high}. Other approaches, such as Linearly Constrained Minimum Variance (LCMV)~\cite{1450747}, extend this by imposing multiple linear constraints to shape the beam pattern. 

Furthermore, the concept of signal sparsity plays a pivotal role in modern array processing~\cite{1614066,1580791,1468495,yang2018sparse}. In the spatial domain, sources are typically sparse—meaning the number of active sources is significantly smaller than the number of possible angular grids. This sparsity allows for the formulation of DoA estimation as a sparse signal recovery problem, where the goal is to reconstruct a sparse source vector from compressed array measurements.

Based on the idea of spatial filtering and the sparsity of the signal in space, We defined an optimal spatial filter $\bm{B^*}$ that focuses on the target signal while suppressing noise and interference, enabling accurate DoA estimation. under the assumptions of narrowband far-field signals and fewer sources than array elements, we show the existence of an optimal spatial filter. Our optimization objective is to find the optimal spatial filter.

Inspired by deep learning, traditional beamforming, and the optimal spatial filter, we propose a deep learning-based beamforming method to approximate the optimal spatial filter $\bm{B^*}$ and apply it to received array signals for target focusing and noise suppression, enabling accurate and robust DoA estimation with interpretability. Under the established existence of $\bm{B^*}$, the universal approximation capability of deep networks~\cite{cybenko1989approximation,hornik1989multilayer,leshno1993multilayer,kidger2020universal,yarotsky2022universal} and extensive validation experiments, our method is theoretically grounded and experimentally validated, providing an interpretable foundation and clarifying the theoretical upper bound. Our main contributions are as follows:

\begin{itemize}
\item \textbf{BeamformNet}. A novel beamforming-based deep learning framework with robust performance under harsh conditions, such as limited snapshots and coherent sources.
\item \textbf{Theoretically grounded and experimentally validated}. The proposed learning paradigm is theoretically grounded and experimentally validated to approximate the optimal spatial filter via implicit spatial signal focusing and noise suppression, and the design of BeamformNet is based on beamforming principles.
\item \textbf{State-of-the-Art Performance}. BeamformNet outperforms existing methods under identical experimental conditions across a wide range of scenarios.
\item \textbf{Open Source}. Code is publicly available on GitHub to foster community advancement.
\end{itemize}

\section{Related work}
\subsection{Deep Learning-Based DoA Estimation}
Over the past decade, numerous neural network-based methods have been proposed for DoA estimation, employing architectures such as MLPs~\cite{chen2020deep,cong2020robust}, CNNs~\cite{zhu2019deep,papageorgiou2021deep,lee2022deep,de2022resnet,shmuel2024subspacenet,gast2025dcd}, RNNs~\cite{merkofer2022deep,merkofer2023music}, and Transformers~\cite{ji2024transmusic}. These approaches formulate DoA estimation either as regression~\cite{merkofer2022deep,merkofer2023music,shmuel2024subspacenet,gast2025dcd} or angular-grid classification~\cite{xu2025deep}, and fall into three categories: pure data-driven models~\cite{papageorgiou2021deep}, model-driven-inspired networks~\cite{merkofer2023music,xu2025deep}, and neural-enhanced model-driven methods~\cite{shmuel2024subspacenet}. Our method adopts an RNN architecture, with the overall model design inspired by model-driven beamforming principles, and frames DoA estimation as a classification task, placing it in the model-driven–inspired category. Crucially, unlike prior works, we propose a novel beamforming-based deep learning framework.

\begin{figure*}[t] 
	\centering
	\includegraphics[width=6in]{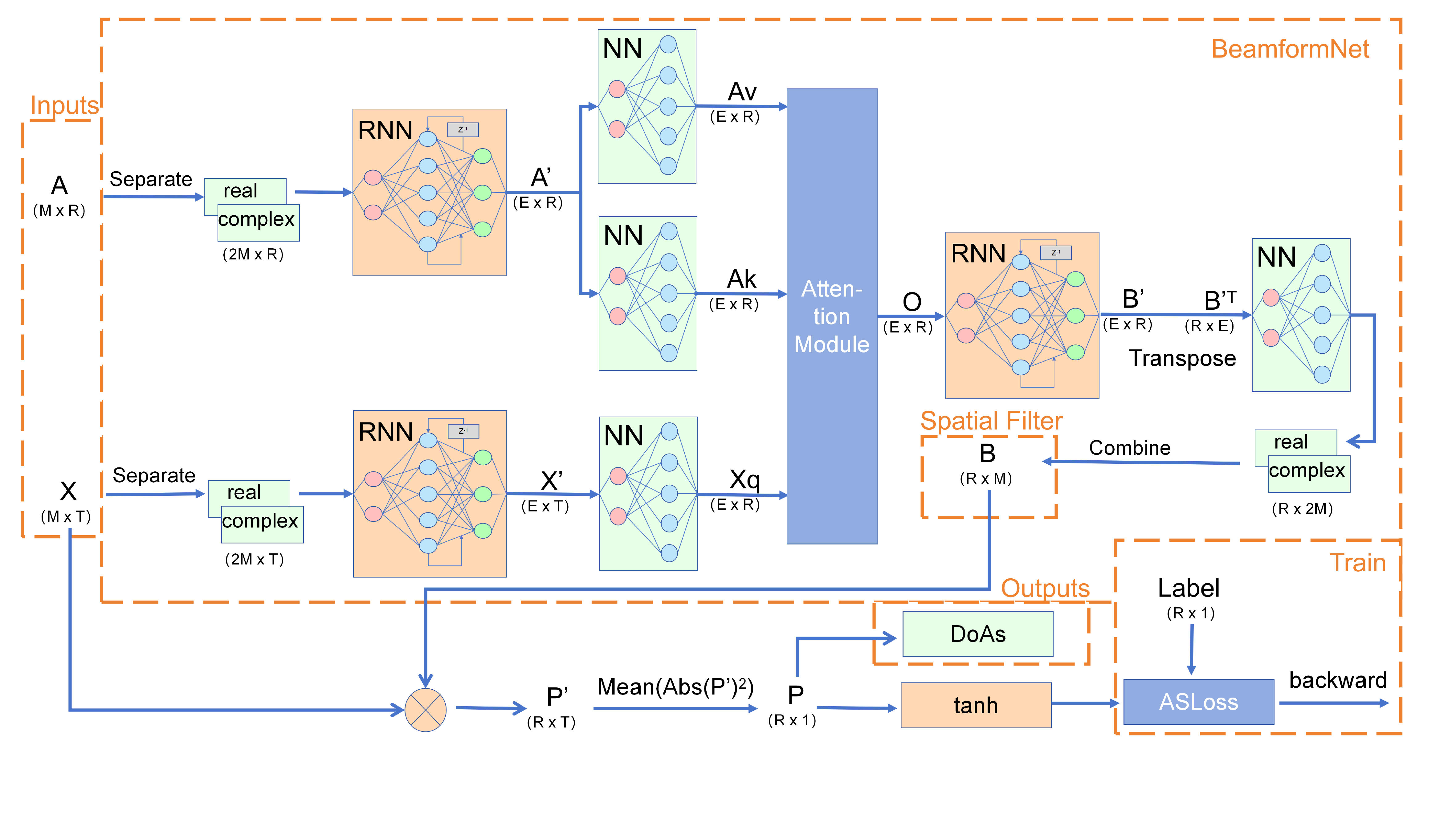}
    \vspace{-20pt}
	\caption{The overall structure of BeamformNet. The proposed learning paradigm is theoretically grounded and experimentally validated to approximate the optimal spatial filter via implicit spatial signal focusing and noise suppression, and the design of BeamformNet is based on beamforming principles.}
	\label{fig:BeamformNet}
\end{figure*}

\subsection{Beamforming}
Beamforming-based DoA estimation methods fall into static~\cite{krim1996two,capon2005high} and dynamic~\cite{zheng2023interpretable,deng2025spatial} categories: the former uses fixed closed-form expression filters, while the latter employs iterative optimization. They can also be classified as non-adaptive~\cite{krim1996two} or adaptive~\cite{capon2005high,deng2025spatial}, depending on whether filter weights are adjusted based on received signals. In terms of implementation methodology, existing approaches can be divided into model-driven methods~\cite{krim1996two,capon2005high,deng2025spatial} and neural network-based methods~\cite{al2022review,zheng2023interpretable,hamza2024sparse}. In addition, beamforming algorithms can be divided into different methods based on different optimization objectives. CBF aims to maximize the Signal-to-Noise Ratio (SNR) under the assumption of white noise, prioritizing robustness but sacrificing resolution~\cite{krim1996two}. Adaptive methods like MVDR~\cite{capon2005high}, on the other hand, seek to minimize the total output power subject to a distortionless constraint in the look direction. This objective implicitly forces the filter to place nulls in the directions of strong interference, thereby maximizing the Signal-to-Interference-plus-Noise Ratio (SINR). Extensions such as LCMV~\cite{1450747} introduce multiple linear constraints to shape the beam pattern further. Our method is a dynamic, adaptive beamforming approach, whose optimization objective is to find the optimal spatial filter, implemented via deep neural networks: its dynamic nature arises from iterative data training, and its adaptivity comes from generating spatial filters in response to input array signals.

\subsection{Exploiting Signal Sparsity}
The spatial sparsity of signal sources—where signals occupy only a few angular directions among a fine grid—has inspired a class of Sparse Signal Recovery (SSR) methods~\cite{1614066,1580791,1468495,yang2018sparse,huang2025efficient}. These methods typically formulate DoA estimation as an inverse problem regularized by sparsity-inducing norms (e.g., $\ell_1$-norm or $\ell_0$ norm). Algorithms such as Orthogonal Matching Pursuit (OMP)~\cite{258082,342465,1337101} and Sparse Bayesian Learning (SBL)~\cite{2001Sparse} utilize this sparsity prior to achieve super-resolution performance, breaking the Rayleigh limit~\cite{van2002optimum}. While effective, traditional sparse reconstruction methods require iterative numerical optimization (e.g., using CVX~\cite{grant2008cvx} or ADMM~\cite{boyd2011distributed}), which is computationally expensive and difficult to deploy in real-time systems. In this work, we integrate the sparsity prior into the deep learning framework. By designing the network to output a spatial spectrum that matches the sparse distribution of ground-truth sources, BeamformNet effectively learns to perform sparse signal recovery in a forward-pass manner, combining the high resolution of sparsity-based methods with the inference speed of neural networks.

\section{System Model and Preliminaries}
\subsection{Signal Model}
Consider $K$ narrowband far-field signals impinging on an $M$-element array. Under ideal conditions—assuming isotropic, perfectly calibrated sensors without mutual coupling—the received signal at time $t$ is modeled as:
{\small
\begin{equation}
\label{eq:x=as+n}
\begin{bmatrix}
x_1(t) \\
x_2(t) \\
\vdots \\
x_M(t)
\end{bmatrix}=
\begin{bmatrix}
\bm{a}_{1}(\omega_0)...\bm{a}_{K}(\omega_0)
\end{bmatrix}
\begin{bmatrix}
s_1(t) \\
s_2(t) \\
\vdots \\
s_K(t)
\end{bmatrix}+
\begin{bmatrix}
n_1(t) \\
 \\
n_2(t) \\
\vdots \\
n_M(t)
\end{bmatrix}
\end{equation}}
of which
{\small
\begin{equation}
\label{eq:bf(a)_i(w_0)}
\bm{a}_i(\omega_0)=
\begin{bmatrix}
e^{-j\omega_0\tau_{1i}} \\
e^{-j\omega_0\tau_{2i}} \\
\vdots \\
e^{-j\omega_0\tau_{Mi}}
\end{bmatrix}\quad i=1,2,...,K
\end{equation}}
and
{\small
\begin{equation}
    \label{eq:w0=2pif}
    \omega_0 = 2\pi f = \frac{2\pi c}{\lambda},
\end{equation}}where $s_i(t)$ is the $i$-th source signal, $x_j(t)$ is the signal received by the $j$-th sensor, $n_j(t)$ is the additive noise at the $j$-th sensor, $\tau_{ji}$ is the time delay of the $i$-th signal at the $j$-th sensor relative to the reference, and $f$, $c$, and $\lambda$ denote frequency, wave speed, and wavelength, respectively.

The expression for the time delay $\tau_{ji}$ is given by:
{\small
\begin{equation}
    \label{eq:tau_ji_in_3D}
    \tau_{ji}= \frac{1}{c} \left( x_j \cos \theta_i \cos \varphi_i+y_j \sin \theta_i \cos \varphi_i+z_j \sin \varphi_i \right),
\end{equation}}where $(x_j, y_j, z_j)$ are the coordinates of the $j$-th sensor with the reference element at the origin, and $\theta_i$, $\varphi_i$ are the azimuth and elevation angles of the $i$-th signal. we refer the reader to references~\cite{Wang2004Spatial,tuncer2009narrowband,friedlander2009wireless,chen2010introduction} for further details.

Following prior work, we consider a half-wavelength uniform linear array (ULA), where $x_j$, $z_j$, and $\varphi_i$ are all zero. Thus, $\tau_{ji}$ is given by:
{\small\begin{equation}
    \label{eq:tau_ji_in_1D}
    \tau_{ji}= \frac{1}{c} \left( y_j \sin \theta_i  \right).
\end{equation}}

DoA estimation is the inverse of signal reception: given $T$ snapshots from an $M$-element ULA, it aims to accurately recover the azimuth angles $\theta_i$ of $K$ sources, assuming $K < M$ and typically $K$ are unknown.

\subsection{Sparse Signal Model}
Since the number of signals is typically limited, literature~\cite{malioutov2005sparse} leverages the sparsity of sources in the spatial domain to construct a sparse signal representation model. Specifically, the entire angular space is discretized into grids with a resolution of $\delta$. A grid contains a signal only if it includes the true direction of this source.Thus, the signal reception model described in Formula~\eqref{eq:x=as+n} can be reformulated as:
{\small \begin{equation}
\label{eq:x=as+n_complete}
\begin{bmatrix}
x_1(t) \\
x_2(t) \\
\vdots \\
x_M(t)
\end{bmatrix}=
\begin{bmatrix}
\bm{a}_{1}(\omega_0)...\bm{a}_{R}(\omega_0)
\end{bmatrix}
\begin{bmatrix}
s_1(t) \\
s_2(t) \\
\vdots \\
s_R(t)
\end{bmatrix}+
\begin{bmatrix}
n_1(t) \\
 \\
n_2(t) \\
\vdots \\
n_M(t)
\end{bmatrix}
\end{equation}}of which
{\small \begin{equation}
    \label{eq:s_i(t)=0_s_i(t)=s_i(t)}
    \begin{cases}s_i(t)=s_i(t), & i \in Grids~with~signals~present \\s_i(t)=0, & i \in other~grids
    \end{cases}
\end{equation}}where $R$ is the number of grids at resolution $\delta$. The discrete sparse model in \eqref{eq:x=as+n_complete} can then be written in matrix form as:
{\small \begin{equation}
\label{eq:X=A_complete S+N}
\bm{X}=\bm{AS}+\bm{N}.
\end{equation}}

\subsection{Beamforming and the Existence of the Optimal Spatial Filter }
Different beamforming techniques can be viewed as deriving distinct spatial filters $\bm{B}$ based on different objectives and then applying them to the array's received signals $\bm{X}$ to estimate the DoA~\cite{van1988beamforming}, that is:
{\small \begin{equation}
    \label{eq:BX=BAS+BN}
    \bm{BX} = \bm{BAS} + \bm{BN},
\end{equation}}DoA is estimated from the spectral peak of the spatial energy spectrum $\bm{P} = [p_1,...,p_i,..,p_R]^T$, where $p_i = \bm{{b_i}^HXX^Hb_i}$ with $p_i$ is the $i$-th element of $\bm{P}$ and $\bm{b_i}$ is the $i$-th column vector of matrix $\bm{B}^H$. For example, in the MVDR algorithm, it provides an analytical solution:
{\small \begin{equation}
    \bm{b_i} = \frac{(\bm{XX}^H)^{-1}\bm{a_i}(w_0)}{\bm{a_i}(w_0)^H((\bm{XX}^H)^{-1})^H\bm{a_i}(w_0)},
\end{equation}}
and in the CBF algorithm,
{\small \begin{equation}
    \bm{b_i} = \frac{\bm{a_i}(w_0)}{M}.
\end{equation}}

We define the optimal spatial filter in a sparse signal model, stating that a filter satisfying the following relation is referred to as the optimal spatial filter:
{\small \begin{equation}
    \label{eq:BX=S}
    \bm{BX = S}.
\end{equation}}

A sufficient condition that satisfies this definition is:
\begin{numcases}{}
\bm{BAS} = \bm{S} \label{eq:BAS=S} \\
\bm{BN} = \bm{0} \label{eq:BN=0}
\end{numcases}with the Formula~\eqref{eq:BAS=S} and Formula~\eqref{eq:BN=0} representing spatial signal focusing and noise elimination, respectively.

Based on this sufficient condition, we can discuss the existence of the optimal spatial filter. Under the assumption $K<M$ and the sparse signal model in Formula~\eqref{eq:x=as+n_complete}, since the degrees of freedom of $\bm{b_i}$ are $M$, while the constraints imposed by $\bm{S}$ and $\bm{N}$ on the degrees of freedom are $K$ and $1$, respectively, there must exist a matrix $\bm{B}$ that simultaneously satisfies Formula~\eqref{eq:BAS=S} and Formula~\eqref{eq:BN=0}.

If the optimal spatial filter is obtained, the spatial energy spectrum $\bm{P}$ of the signal can be calculated. That is, we can accurately estimate the DoA based on the definition in Formula~\eqref{eq:BX=S} with $p_i = \bm{{b_i}^HXX^Hb_i} =  ||s_i(t)||^2$ and the sparsity described in Formula~\eqref{eq:s_i(t)=0_s_i(t)=s_i(t)}. 

The literature~\cite{deng2025spatial} provides a model-driven method for solving this optimal spatial filter under the condition of a suitable two-dimensional array. But this method is effective only under certain array conditions. We propose using a neural network to solve for the optimal spatial filter $\bm{B^*}$, which overcomes constraints, successfully applies to the ULA array, and achieves better performance, as shown in experiments.

\subsection{Approximate the Optimal Spatial Filter by Neural Network}
The Universal Approximation Theorem serves as the theoretical foundation of deep learning, stating that under mild conditions, neural networks can approximate any mapping from multiple independent variables to fewer dependent variables with arbitrary precision~\cite{cybenko1989approximation,hornik1989multilayer,leshno1993multilayer,kidger2020universal,yarotsky2022universal}. Therefore, for each element within the existing optimal filter $\bm{B^*}$, it is theoretically feasible to construct a deep neural network that approximately maps the information to that specific element, i.e.,
{\small \begin{equation}
    f(information) = \bm{B^*}(i,j),
\end{equation}}where $f$ denotes the mapping function acted upon by neural networks, and $\bm{B^*}(i,j)$ represents any element within the optimal filter $\bm{B^*}$. The degree of approximation to $\bm{B^*}$ depends on the network structure and design.

In our proposed BeamformNet method, we construct the mapping function $f$ based on RNNs, which approximately maps the information—including the steering vector $\bm{A}$ and the received data matrix $\bm{X}$—to the optimal filter $\bm{B^*}$. As BeamformNet is designed based on beamforming principles (see the~\ref{sec:BeamformNet}.BeamformNet section) and experimentally validated (see the~\ref{sec:results}.Results section), it can effectively approximate the optimal spatial filter across a wide range of scenarios.

\section{BeamformNet}
\label{sec:BeamformNet}

\begin{figure}[t] 
	\centering
	\includegraphics[width=3.5in]{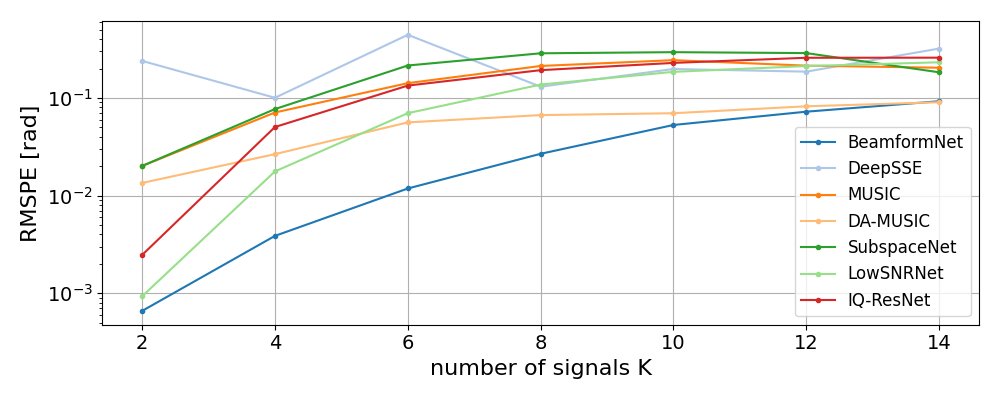}
	\caption{RMSPE of DoA estimation with different $K$.}
	\label{fig:K}
\end{figure}

\begin{figure}[t] 
	\centering
	\includegraphics[width=3.5in]{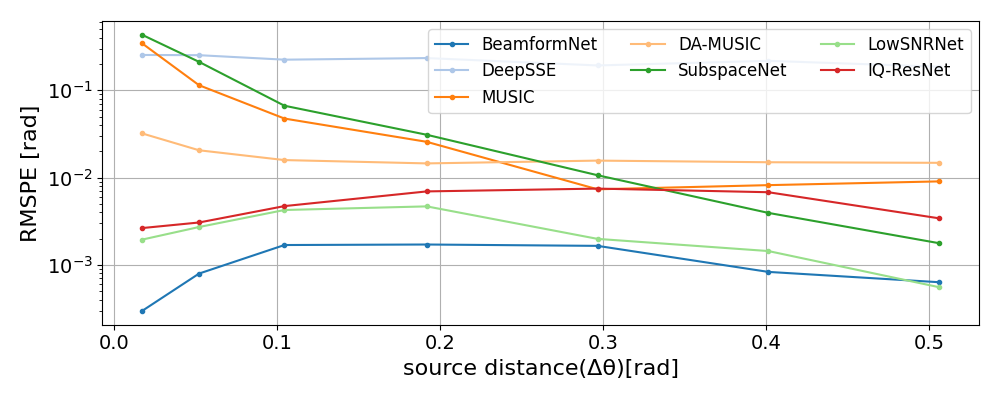}
	\caption{RMSPE of DoA estimation with different $\Delta\theta$ [rad].}
	\label{fig:Delta_theta}
\end{figure}

\begin{figure}[t] 
	\centering
	\includegraphics[width=3.5in]{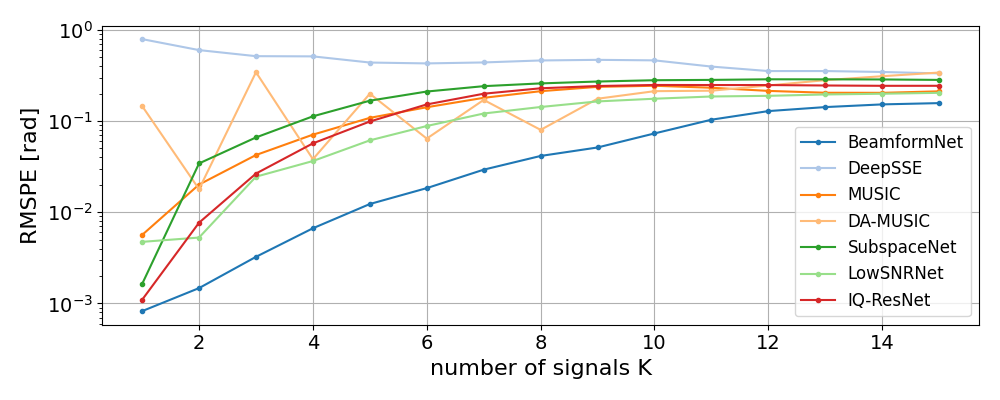}
	\caption{The generalization of each algorithm with respect to $K$.}
	\label{fig:K_robust}
\end{figure}

\begin{figure}[t] 
	\centering
	\includegraphics[width=3.5in]{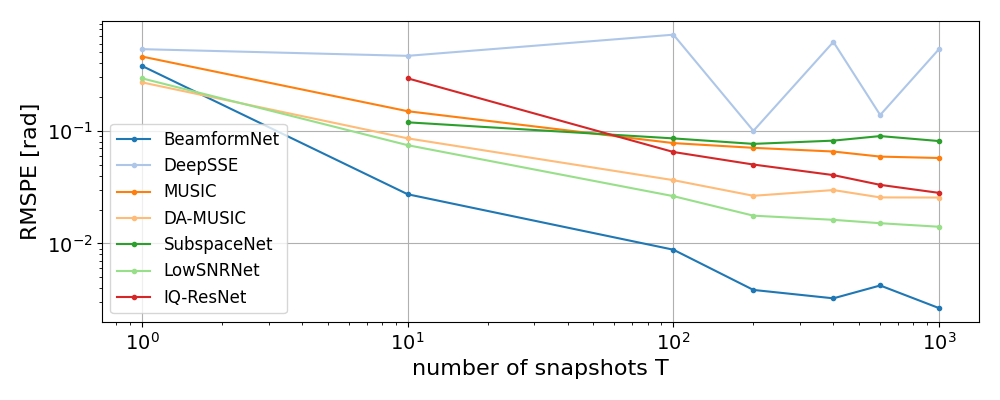}
	\caption{RMSPE of DoA estimation with different T.}
	\label{fig:T}
\end{figure}

\begin{figure}[t] 
	\centering
	\includegraphics[width=3.5in]{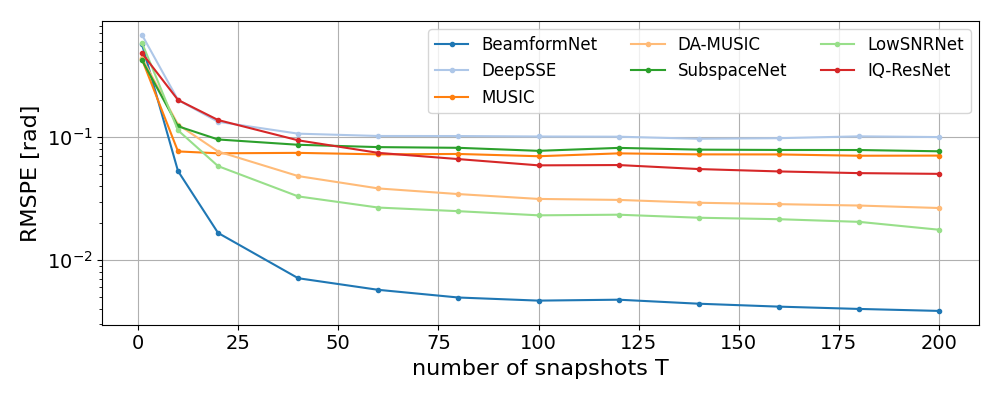}
	\caption{The generalization of each algorithm with respect to $T$.}
	\label{fig:T_robust}
\end{figure}

\begin{figure}[t] 
	\centering
	\includegraphics[width=3.5in]{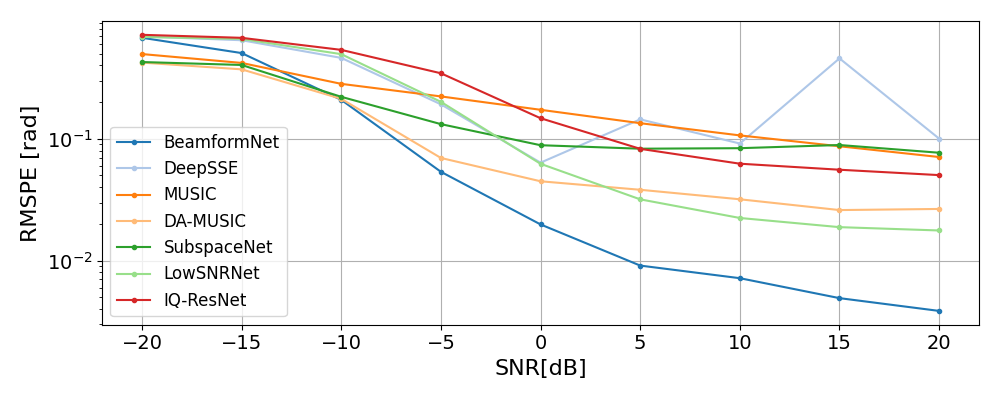}
	\caption{RMSPE of DoA estimation with different SNRs.}
	\label{fig:SNR}
\end{figure}

\begin{figure}[t] 
	\centering
	\includegraphics[width=3.5in]{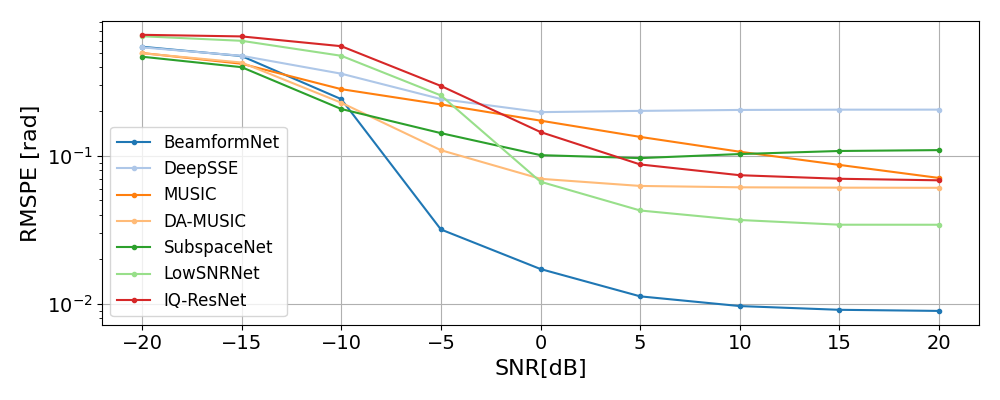}
	\caption{The generalization of algorithms with respect to SNR.}
	\label{fig:SNR_robust}
\end{figure}

\begin{figure}[t] 
	\centering
	\includegraphics[width=3.5in]{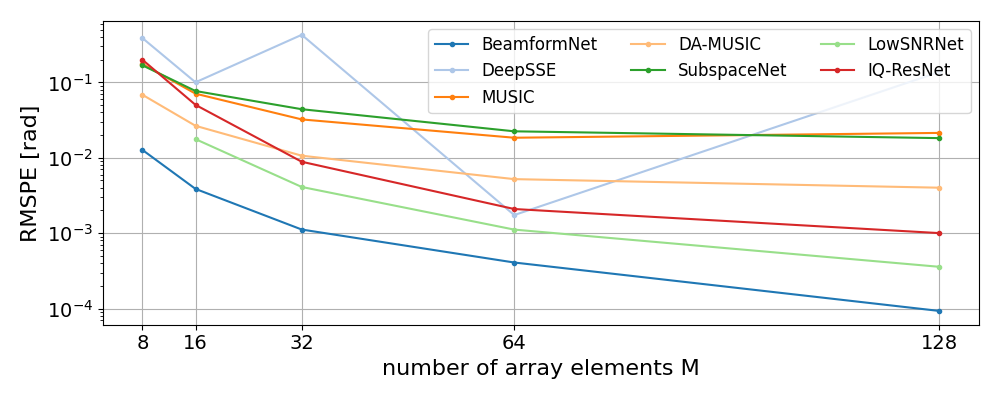}
	\caption{RMSPE of DoA estimation with different M.}
	\label{fig:M}
\end{figure}

\begin{figure}[t] 
	\centering
	\includegraphics[width=3.5in]{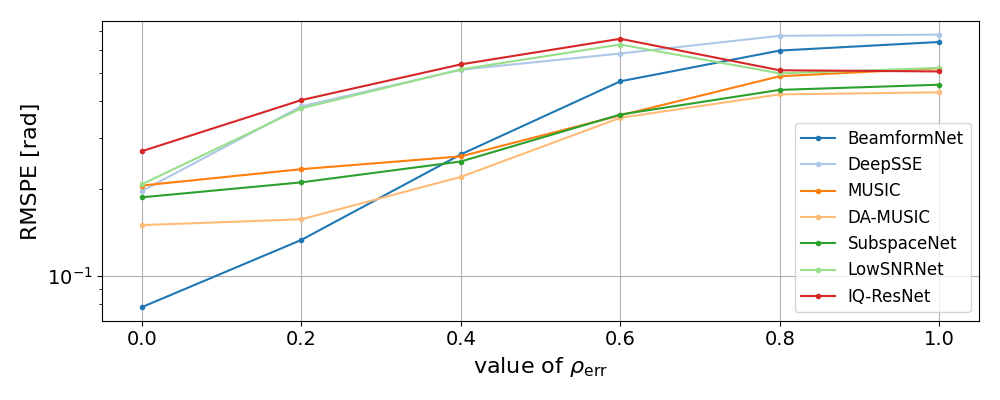}
	\caption{RMSPE of DoA estimation with different mismatch.}
	\label{fig:p_err}
\end{figure}

\begin{figure}[t] 
	\centering
	\includegraphics[width=3.5in]{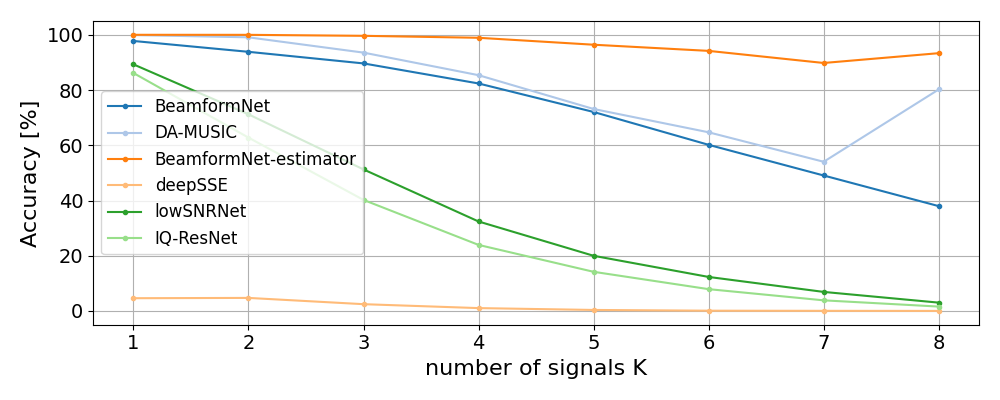}
	\caption{Accuracy under unknown number of signals $K$.}
	\label{fig:accuracy}
\end{figure}

\begin{table}[t]
\caption{The number [Million] of parameters variation of different methods with increasing number of array elements $M$.}
\label{tb:parameters}
\begin{center}
\begin{small}
\begin{sc}
\begin{tabular}{@{}p{2cm}p{0.45cm}p{0.45cm}p{0.675cm}p{0.675cm}p{0.675cm}p{0.675cm}@{}}
\toprule
Method/$M$      & 8 & 16 & 32 & 64 & 128 & 256 \\ \midrule
LowSNRNet   &  ---   &   28.25   &  1.62e2    &   8.33e2   &   3.78e3 & 1.61e4   \\
IQResNet    &  1.85   &   1.86   &  1.87   &  1.90    &  1.93 &2.01    \\
DeepSSE     &  0.57   &    0.57  &   0.58  &  0.59    &    0.62    & 0.66\\
SubspaceNet &   0.04  &   0.04   &  0.04    &  0.04    &    0.04   &0.04\\
DA-MUSIC    &  0.01   &   0.04   &   0.19   &  1.28    &  9.26      & 70.48\\
BeamformNet &  6.60   &   6.66   &   6.77   &   7.00   & 7.46  & 8.38     \\ \bottomrule
\end{tabular}
\end{sc}
\end{small}
\end{center}
\end{table}

\begin{table*}[t]
\caption{Coherent and incoherent sources with different methods.}
\label{tb:coherent}
\begin{center}
\begin{small}
\begin{sc}
\begin{tabular}{@{}llllllll@{}}
\toprule
Sources      & BeamformNet & LowSNRNet & IQResNet & DeepSSE & subspaceNet & DA-MUSIC & MUSIC  \\ \midrule
Coherent   &  \bf{0.125556}   &  0.262917    &   0.350655   &   0.571993   &   0.255021 &  0.126696 & 0.329191  \\
Incoherent    &  \bf{0.078276}   &  0.207183    &   0.269299   &  0.196394    &  0.186812 & 0.149886 &   0.205061  \\ \bottomrule
\end{tabular}
\end{sc}
\end{small}
\end{center}
\end{table*}

\begin{table*}[t]
\caption{Real-world speech acoustic source localization with different methods. BB-MUSIC is broadband extension of MUSIC using
independent frequency bins (IFBs). BB-MUSIC is implemented using 10 Hz per IFB form 100 Hz to 1000 Hz;}
\label{tb:realman}
\begin{center}
\begin{small}
\begin{sc}
\begin{tabular}{@{}llllllll@{}}
\toprule
      & BeamformNet & LowSNRNet & IQResNet & DeepSSE & subspaceNet & DA-MUSIC & BB-MUSIC \\ \midrule
RealMAN   &  \bf{0.179377}   &  0.354476    &   0.23065   &   0.771702   &  0.618011 & 0.438992 & 0.435888     \\\bottomrule
\end{tabular}
\end{sc}
\end{small}
\end{center}
\end{table*}

\subsection{Context Extraction in Time and Spatial Domains}
 We propose BeamformNet, whose overall structure is illustrated in Figure~\ref{fig:BeamformNet}, to solve for the optimal filter $\bm{B^*}$. Inspired by model-driven adaptive beamforming, BeamformNet takes the array manifold matrix $\bm{A}$, which contains all grid azimuth angles, and the array received signal $\bm{X}$ as inputs. Inputs $\bm{X}$ and $\bm{A}$ represent $M \times T$ time series information and $M \times R$ spatial series information, respectively. By taking $\bm{X}$ and $\bm{A}$ as inputs, BeamformNet adjusts the array response to match varying signal characteristics and adaptively generates the optimal spatial filter.

 To preserve the phase, BeamformNet splits $\bm{A}$ and $\bm{X}$ into real and imaginary parts, concatenates them, and feeds each into separate bidirectional RNNs to extract contextual features in  the time and spatial domains in order to adaptively generate the spatial filter $\bm{B}$. Some works~\cite{zhou2018direction,su2024co,ma2024doa} show that virtual array elements can enhance array element dimensions for better DoA estimation. Accordingly, we expand the array element dimensions of $\bm{A}$ and $\bm{X}$ from $2M$ to $E$ ($E > 2M$) after passing through the RNN, yielding an enriched representation that captures both temporal and spatial contexts with additional array element dimensions:
\begin{numcases}{}
    \bm{A'} = RNN(concat(separate(\bm{A}))) \\
    \bm{X'} = RNN(concat(separate(\bm{X})))
\end{numcases}with 
{\small \begin{equation}
\begin{gathered}
    \Im(\cdot),~\Re(\cdot) = separate(\cdot)\\
    [\Im(\cdot),~\Re(\cdot)]^T = concat(\Im(\cdot),\Re(\cdot)),
\end{gathered}
\end{equation}}where $\Im(\cdot)$ and $\Re(\cdot)$ denote the imaginary and real parts of each matrix element, respectively, to form a new matrix.

\subsection{Attention module}
Inspired by traditional model-driven beamforming that fuses $\bm{X}$ and $\bm{A}$, we introduce an attention module to enable BeamformNet to integrate $\bm{A'}$ and $\bm{X'}$. Moreover, since traditional model-driven methods compute spatial filters $\bm{B}$ at each grid point using the corresponding steering vectors in $\bm{A}$, our attention mechanism enables BeamformNet to assign adaptive weights to $\bm{A}$ for different grids, focusing on the target signal’s spatial region.

Before the attention module, BeamformNet uses two-layer linear neural network (NN) to align $\bm{A'}$ and $\bm{X'}$ into the query, key, and value of the attention mechanism:
{\small \begin{equation}
\begin{gathered}
    \bm{A}_v = NN(\bm{A'})\\
    \bm{A}_k = NN(\bm{A'})\\
    \bm{X}_q = NN(\bm{X'}).
\end{gathered}
\end{equation}}Here, $\bm{X}_q$ serves as the query, which retrieves information from the value $\bm{A}_v$ based on the key $\bm{A}_k$. This guides the model to focus on signal-present spatial grids for yielding an adaptive spatial filter B. The attention is computed as:
{\small \begin{equation}
    \bm{O} = softmax(\frac{\bm{X}_q{\bm{A}_k}^T}{\sqrt{R}})\bm{A}_v.
\end{equation}}$\bm{O}$ is the attention module output, fusing spatial and temporal information to better characterize the target signal space.

\subsection{Spatial Filter}
After obtaining $\bm{O}$, BeamformNet uses an RNN to produce a spatial filter $\bm{B'}$ that has a higher element dimension:
{\small \begin{equation}
    \bm{B'} = RNN(\bm{O}).
\end{equation}}Subsequently, BeamformNet employs a two-layer linear network to project $\bm{B'}$ to the actual array element dimension, yielding the real and imaginary parts of the final spatial filter $\bm{B}$, which are then combined:
{\small \begin{equation}
\begin{gathered}
    [\Im(\bm{B}),\Re(\bm{B})]^T = NN(\bm{B'}^T)\\
    \bm{B} = combine([\Im(\bm{B}),\Re(\bm{B})]^T) = \Re(\bm{B}) + j \cdot \Im(\bm{B}).
\end{gathered}
\end{equation}}The spatial filter $\bm{B}$ is then applied to the array received signal $\bm{X}$ to obtain the spatial spectrum $\bm{P'}$ over $T$ snapshots:
{\small \begin{equation}
    \bm{P'} = \bm{BX}.
\end{equation}}Then we calculate the average energy spectrum $\bm{P} = [p_1,\dots,p_i,\dots,p_R]^T$, where $p_i = \frac{\bm{b_i}^H\bm{XX}^H\bm{b_i}}{T}=\frac{||[s_i(1),\dots,s_i(T)]^T||^2}{T}$, under the $T$ snapshots for the spatial spectrum $\bm{p'}$. It is equivalent to performing the following operation on matrix $\bm{P'}$:
{\small \begin{equation}
    \bm{P} = Mean(Abs(\bm{P'})^2).
\end{equation}}$Abs$ denotes element-wise modulus, and $Mean$ denotes averaging over the snapshot dimension.

According to the sparsity in Formula~\eqref{eq:s_i(t)=0_s_i(t)=s_i(t)}, $p_i$ reflects the average signal energy over $T$ snapshots—nonzero at signal-present grids and near-zero elsewhere—so DoAs can be estimated by identifying peaks in $\bm{P}$.

\subsection{Loss function}
After obtaining the spatial energy spectrum $\bm{P}$, one possible approach is to directly learn according to Formula~\eqref{eq:BX=S}, which calculates the error between the obtained spatial energy spectrum $p_i = \bm{b_i^HXX^Hb_i}$ and the sparse spatial signal energy spectrum $||s_i(t)||^2$, such as using MSE (Mean  Squared Error). Because our goal is to solve DoA, we are not interested in energy. So we directly solve for the direction in which the target angle is located. The DoAs obtained by searching the spatial energy spectrum $\bm{P}$ through spectral peaks cannot be directly used to compute gradients for backpropagation. In order to enable BeamformNet to backpropagate gradients for training, we apply the $tanh$ activation function to $\bm{P}$. Since each element in $\bm{P}$ lies in the range $[0,+\infty)$, applying $tanh$ maps the signal energy in each grid cell to a probability distribution over $[0,1]$. Accordingly, BeamformNet formulates model training as a multi-label classification problem, which maximizes the probability of the grid where the target angle is located. 

The above process uses the $tanh$ function (monotonically increasing) to convert $\bm{P}$ into probability and utilizes the sparsity of $\bm{S}$ for classification, which implies that the direction energy of the signal in $\bm{P}$ is higher (signal focusing), while other directions are lower or zero (noise suppression). It is equivalent to letting the network implicitly learn Formulas~\eqref{eq:BAS=S} and~\eqref{eq:BN=0} (see Section~\ref{sec:Intuitive qualitative results} for more details), although our network does not explicitly learn these two formulas. It is worth noting that although we do not directly and strictly constrain the network to adhere to Formula~\eqref{eq:BX=S} (since we impose no constraints on energy), the suppression of noise and separation of signals compel BeamformNet to approximate Formula~\eqref{eq:BX=S}, thereby achieving near-optimal spatial filtering.

As indicated in Formula~\eqref{eq:s_i(t)=0_s_i(t)=s_i(t)}, the spatial sparsity of signals implies that positive labels are relatively scarce. To mitigate the issue of class imbalance and enhance focus on positives, we use Asymmetric Loss (ASL)~\cite{ridnik2021asymmetric}, a loss tailored for imbalanced multi-label classification:
{\small \begin{equation}
    ASL(\bm{\rho},\bm{\hat{\rho}}) = \sum_{i=1}^R\mathcal{L}_{ASL}(\rho_i,~\hat{\rho}_i),
\end{equation}}where
{\small \begin{equation}
\mathcal{L}_{ASL}(\rho_{i},\hat{\rho}_{i})= 
\begin{cases}
-(1-\hat{\rho}_{i})^{\gamma_{+}}\log(\hat{\rho}_{i}), & \text{for } \rho_{i}=1, \\
-(\rho_{m})^{\gamma_{-}}\log(1-\rho_{m}), & \text{for } \rho_{i}=0.
\end{cases}
\end{equation}}Here, $\rho_i \in \{0,1\}$ is the ground-truth label for the $i$-th angular grid, $\hat{\rho}_{i} \in [0,1]$ is the model’s predicted confidence, and $\rho_m = max(\hat{\rho}_{i}-\eta_m,0)$. The focusing parameters $\gamma_{+} =1$ and $\gamma_{-} = 4$ emphasize positive and negative grids, respectively, with $\eta_m=0.05$, following~\cite{ridnik2021asymmetric}.

\section{Experiments Settings}
\subsection{Dataset}

We follow~\cite{merkofer2023music}’s data generation method to create our dataset, simulating measurements $\bm{X}(t)$ per Formula~\eqref{eq:x=as+n}. The $K$ signals $s_i(t)$ and $M$ noise terms $n_i(t)$ are independently drawn from the complex Gaussian distribution $\mathcal{CN}(0,1)$ across all $T$ snapshots, modeling random amplitudes and phases, with noise scaled to maintain a fixed Signal-to-Noise Ratio (SNR). We use a Uniform Linear Array (ULA) with center frequency $f$=1000 Hz, propagation speed $c$=340 m/s, and half-wavelength element spacing. $K$ signal azimuths are independently sampled from a uniform distribution $\mathcal{U}(-\pi/2,\pi/2)$. The grid resolution is set to $\delta=1^\circ$. To avoid overfitting, we randomly generate 90k training samples per epoch and 10k validation/test samples each, all using distinct random seeds.

\subsection{Training}
During the experiments, we compared different methods, including DA-MUSIC~\cite{merkofer2023music}, DeepSSE~\cite{xu2025deep}, SubspaceNet~\cite{shmuel2024subspacenet}, IQResNet~\cite{zheng2024deep}, LowSNRNet~\cite{papageorgiou2021deep} and classic MUSIC~\cite{schmidt1986multiple}. For DA-MUSIC, we manually provide a perfect source number estimator (100\% accuracy). For BeamformNet, the hyperparameter $E$=256 (See the subsection~\ref{sec:E} for the ablation experiment of E). All methods used identical experimental settings. We trained with a batch size of 32, a learning rate of 0.0001, and the Adam optimizer~\cite{kingma2014adam} ($\beta_1=0.9$, $\beta_2=0.999$) for up to 100 epochs, using early stopping if the validation performance showed no improvement for 20 epochs. In validation, DA-MUSIC and SubspaceNet, being regression-based, were evaluated with RMSPE~\cite{Routtenberg2012Bayesian,Routtenberg2013Non}, while BeamformNet and other classification-based methods used the F1 score.

\subsection{Evaluation}
Final DoA estimates follow each method’s standard procedure. For BeamformNet, while it theoretically computes an optimal spatial filter $\bm{B^*}$, practical constraints result in an approximation of this filter, expressed as:
\begin{numcases}{}
\bm{BAS} \approx \bm{S} \label{BAS approx S} \\
\bm{BN} \approx \bm{0} 
\end{numcases}The approximations indicate that the resulting spatial power spectrum $\bm{P}$ contains weak spatial noise and signal interference. To filter them out, we threshold the probabilities $\rho_i=tanh(p_ 
i)$ at 0.5, discarding grids with $\rho_i<0.5$. The final DoAs are then obtained by peak search on the filtered $\bm{P}$. Following prior studies, the RMSPE is adopted as the final evaluation metric for all compared methods. It can be expressed as:
{\small \begin{equation}
    \operatorname{RMSPE}(\bm{\theta},\bm{\hat{\theta}}) = \min_{\bm{Z} \in \mathcal{Z}_{K}} 
    \left( \frac{1}{K} \left\| \bmod_{\pi} (\boldsymbol{\theta} - \bm{Z} \hat{\boldsymbol{\theta}}) \right\|^{2} \right)^{\frac{1}{2}},
\end{equation}}where $\mathcal{Z}_{K}$ is the set of all $K \times K$ permutations of $\bm{\hat{\theta}}$ and $\bmod_{\pi}$ denotes the modulo operation with respect to $\pi$. Since RMSPE requires equal numbers of estimated and true DoAs, we follow~\cite{papageorgiou2021deep}: Based on the true number of signals $K$ in the label, we truncate the DoAs according to their energy levels if the estimated number exceeds $K$, or pad with the top-$K$ highest energy grid's azimuth angle if the estimated number is less than $K$.

\section{Results}
\label{sec:results}
\subsection{Number of Signals $K$}
The greater the number of signal sources, the more complex the received signals become. To assess the impact of the number of signals $K$ on BeamformNet, we conducted experiments under the settings of $M=16$, $T=200$, $SNR=20 dB$, and incoherent signal sources. As shown in Figure~\ref{fig:K}, BeamformNet significantly outperforms all baselines, especially the model-driven MUSIC algorithm.

To evaluate algorithmic resolution, Figure~\ref{fig:Delta_theta} presents the RMSPE for localizing two closely spaced, non-coherent sources separated by an angular distance $\Delta\theta$. As $\Delta\theta$ approaches approximately 0.05 radians, the MUSIC algorithm fails to resolve them, whereas BeamformNet maintains consistently low localization error across all $\Delta\theta$, demonstrating superior resolution relative to competing methods.

In addition, we evaluated generalization with respect to the signal number $K$, training on $K\in\{2,4,6,8\}$ with $M=16$, $T=200$, $SNR = 20 dB$, and incoherent sources. As shown in Figure~\ref{fig:K_robust}, BeamformNet generalizes well—even to unseen $K$—outperforming all methods, while DA-MUSIC shows a fluctuating performance pattern and poor generalization on unseen source counts.

\subsection{Number of Snapshots $T$}
Fewer snapshots $T$ mean less information in the received signals, making DoA estimation more challenging. Therefore, we investigated the impact of the snapshot number $T$ with $M=16$, $SNR = 20 dB$, $K=4$, and incoherent sources. As shown in Figure~\ref{fig:T}, in a single snapshot setup, DA-MUSIC achieves the best DoA estimation performance, whereas BeamformNet's performance is lower than that of LowSNRNet. We attribute this to the RNN’s limited ability to extract temporal information from a single snapshot, leading to its weaker performance. However, as $T$ increases, the RNN gradually captures more useful temporal features, enabling BeamformNet to exhibit optimal performance under non-single snapshot conditions.

We further evaluated the generalization capability of each algorithm concerning the number of snapshots $T$ by testing models previously trained at $T=200$ across $T \in [1,200]$ with replicated padding inputs to match the training length. As shown in Figure~\ref{fig:T_robust}, BeamformNet achieves the best performance under non-single-snapshot conditions and demonstrates a strong generalization ability across $T$.

\subsection{Signal-to-Noise Ratio}
The SNR measures signal strength relative to noise. A higher SNR indicates better signal quality and more accurate information transmission. We studied SNR's impact on BeamformNet under $M=16$, $K=4$, $T=200$, and incoherent sources, comparing it with other methods. Figure~\ref{fig:SNR} shows that DA-MUSIC performs best at low SNRs (-20 dB and -15 dB), while SubspaceNet and BeamformNet show comparable results. When the SNR exceeds -15 dB, BeamformNet outperforms all other methods, especially at high SNRs. We attribute BeamformNet's weaker performance under extremely low SNRs to the fact that, although it suppresses noise at low SNRs (as shown in Table~\ref{tb:noise_suppression}), the noisy array received signal $\bm{X}$ impacts BeamformNet's ability to separate and focus signals (Fomula~\ref{BAS approx S}). Enhancing BeamformNet's signal separation and focusing performance under extremely low SNRs represents a promising area for future research. Nevertheless, overall, BeamformNet displays strong performance across the range of SNR conditions.

We also investigated the generalization capability of BeamformNet with respect to the SNR by training models at $SNR \in \{-10dB, 0dB, 10dB\}$ with $M = 16$, $T = 200$, $d = 4$, and incoherent sources. As shown in Figure~\ref{fig:SNR_robust}, SubspaceNet generalizes best below -10 dB, while at SNRs above -10 dB, BeamformNet demonstrates superior generalization capability across varying SNR levels compared to other methods.

\subsection{Number of Array Elements $M$}
Generally, the more array elements $M$ there are in an array, the stronger the focusing capability of the array becomes. To investigate whether the performance of the proposed method improves with more array elements, We studied the effect of array elements $M$ on performance with $K=4, SNR = 20 dB, T=200$, and incoherent sources. Figure~\ref{fig:M} shows that BeamformNet consistently outperforms other methods across all $M$, with performance improving as $M$ increases. Notably, as Table~\ref{tb:parameters} shows, BeamformNet’s parameter count does not increase dramatically like those in LowSNRNet and DA-MUSIC as $M$ grows. Instead, BeamformNet maintains an appropriate number of parameters while achieving optimal performance, making it well-suited for scaling up to large-scale arrays with many elements.

We also evaluated BeamformNet under array mismatch due to sensor position errors in the 2D horizontal plane, modeled as in Formula~\eqref{eq:x=as+n} and~\eqref{eq:tau_ji_in_3D} with $z_j = 0$ and $\varphi_i = 0$. In the presence of such positional errors, the actual steering vector can be expressed as:
{\small \begin{equation}
\mathbf{a}_i(\omega_0) = \exp^{-j\frac{\omega_0}{c} \left[ (\bm{x}^T + \Delta \bm{x}^T)\cos\theta_i + (\bm{y}^T + \Delta \bm{y}^T)\sin\theta_i \right]},
\end{equation}}where $\bm{x}, \bm{y} \in \mathbb{R}^{1 \times M}$ are the ideal sensor positions, and $\Delta\bm{x}, \Delta\bm{y}$ are perturbations with entries drawn independently from $\mathcal{U}(-\rho_{\text{err}}\times\frac{\lambda}{2}, \rho_{\text{err}}\times\frac{\lambda}{2})$. $\frac{\lambda}{2}$ represents half wavelength, which is the element spacing of ULA. The mismatch severity is controlled by $\rho_{\text{err}}$. As shown in Figure~\ref{fig:p_err}, BeamformNet outperforms all methods under small mismatches ($\rho_{\text{err}} < 0.4$) but performs poorly under large positional errors.

\subsection{Coherent and Incoherent Sources}
For many traditional model-driven methods, coherent signals violate the fundamental assumption of statistical independence among sources, leading to rank deficiency and consequently causing these methods to fail. To evaluate the performance of BeamformNet under both coherent and incoherent signal conditions, we conducted experiments with $K \in \{2, 4, 6, 8\}$, $SNR\in\{-10dB, -5dB, 0dB, 5dB, 10dB\}$, $M = 16$, and $T = 200$. As shown in Table~\ref{tb:coherent}, BeamformNet consistently achieves the best performance regardless of whether the sources are coherent or non-coherent.

\begin{figure}[t] 
	\centering
	\includegraphics[width=2.5in]{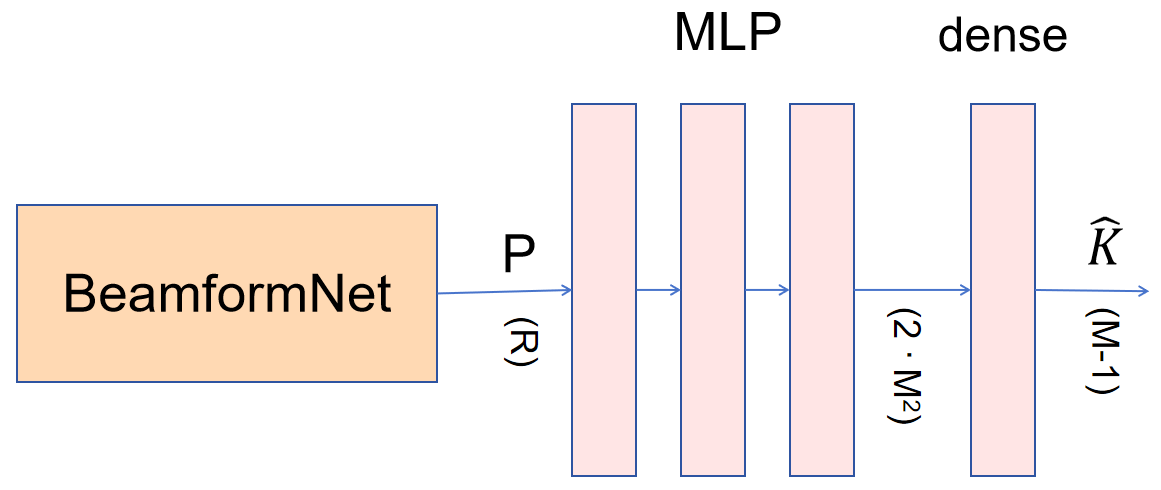}
	\caption{The structure of BeamformNet-estimator.}
	\label{fig:BeamformNet_estimator}
\end{figure}

\begin{figure}[t] 
	\centering
	\includegraphics[width=3.5in]{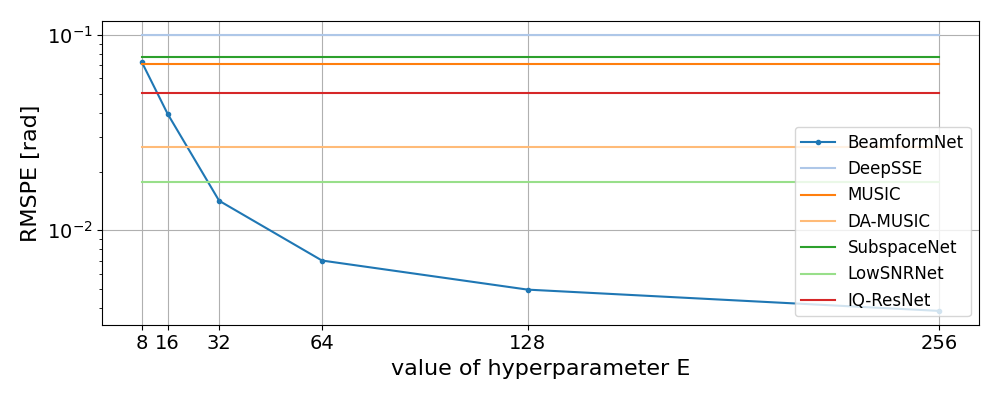}
	\caption{RMSPE of DoA Estimation with different value of $E$.}
	\label{fig:E}
\end{figure}

\begin{table}[t]
\caption{The number [Million] of parameters variation of BeamformNet with increasing value of hyperparameter $E$.}
\label{tb:E}
\begin{center}
\begin{small}
\begin{sc}
\begin{tabular}{@{}p{2cm}p{0.45cm}p{0.45cm}p{0.675cm}p{0.675cm}p{0.675cm}p{0.675cm}@{}}
\toprule
Method/$E$      & 8 & 16 & 32 & 64 & 128 & 256 \\ \midrule
BeamformNet &  0.24  &   0.26   &   0.35   &   0.66   & 1.87  & 6.66     \\ \bottomrule
\end{tabular}
\end{sc}
\end{small}
\end{center}
\end{table}

\subsection{Accuracy under Unknown Number of Signals $K$}
Generally, the $K$ is unknown, requiring algorithms to accurately estimate the source count. To investigate the estimation accuracy of BeamformNet, we compared the accuracy of different methods under varying numbers of sources in the experimental setting of $K = \{1, 2, 3, 4, 5, 6, 7, 8\}$, $SNR = 0 dB$, $M = 16$, and $T = 200$. As shown in Figure~\ref{fig:accuracy}, BeamformNet and DA-MUSIC achieve comparable accuracy, outperforming other methods. However, DA-MUSIC separately employs an MLP as a source number estimator during the estimation process, whereas BeamformNet obtains the source number merely by identifying spectral peaks. For a fair comparison, we followed the approach of DA-MUSIC and equipped BeamformNet with a separate, dedicated MLP-based source number estimator. BeamformNet and DA-MUSIC maintain the same network structure for estimating the number of signal sources, with the only difference being that DA-MUSIC takes eigenvalues as input while BeamformNet takes the spatial energy spectrum $\bm{P}$ as input. The structure of BeamformNet-estimator is shown in Figure~\ref{fig:BeamformNet_estimator}. As illustrated in Figure~\ref{fig:accuracy}, the MLP-based source number estimator BeamformNet-estimator achieves optimal performance across all source numbers, with an average accuracy of 96.54\%, significantly surpassing that of DA-MUSIC (81.25\%).

\begin{figure*}[t]
    \centering
    \begin{subfigure}{0.45\linewidth}
        \centering
        \includegraphics[width=\linewidth]{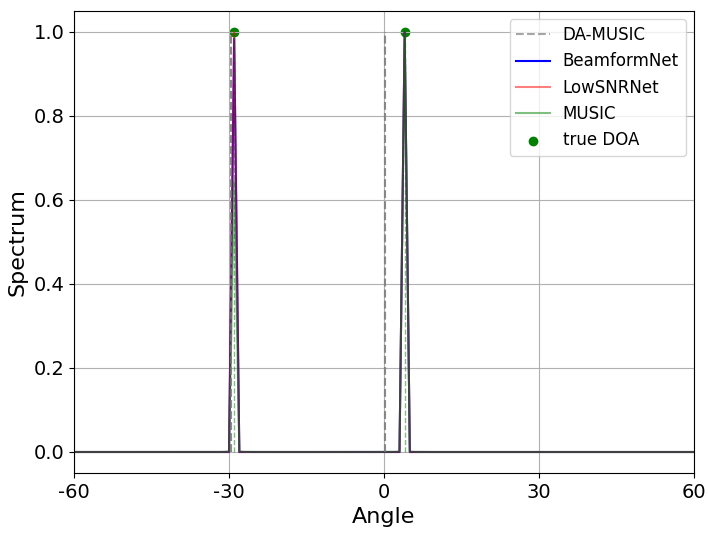}
        \caption{$K=2$}
        \label{fig:sub1}
    \end{subfigure}
    \hfill
    \begin{subfigure}{0.45\linewidth}
        \centering
        \includegraphics[width=\linewidth]{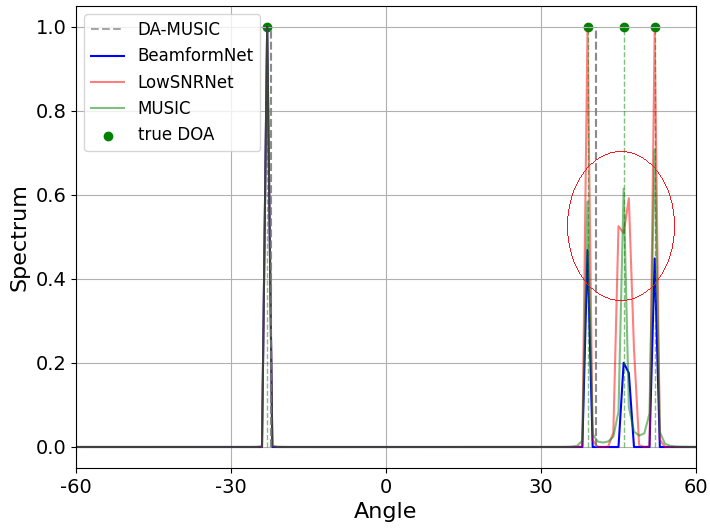}
        \caption{$K=4$}
        \label{fig:sub1}
    \end{subfigure}
    \vfill
    \begin{subfigure}{0.45\linewidth}
        \centering
        \includegraphics[width=\linewidth]{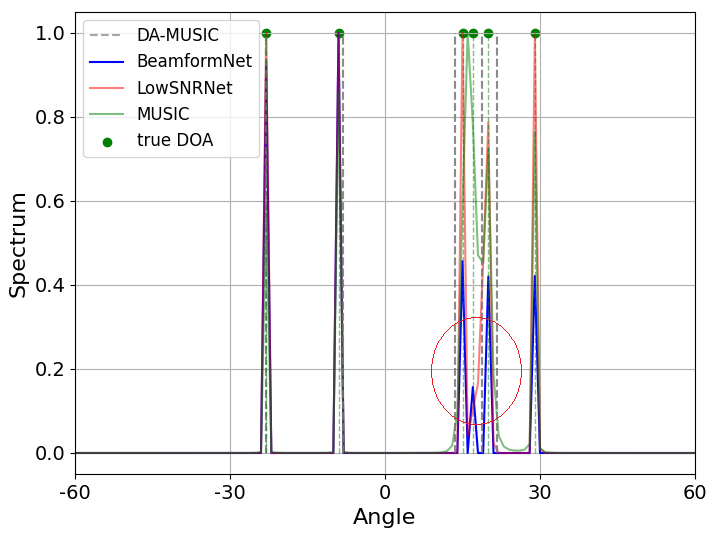}
        \caption{$K=6$}
        \label{fig:sub1}
    \end{subfigure}
    \hfill
    \begin{subfigure}{0.45\linewidth}
        \centering
        \includegraphics[width=\linewidth]{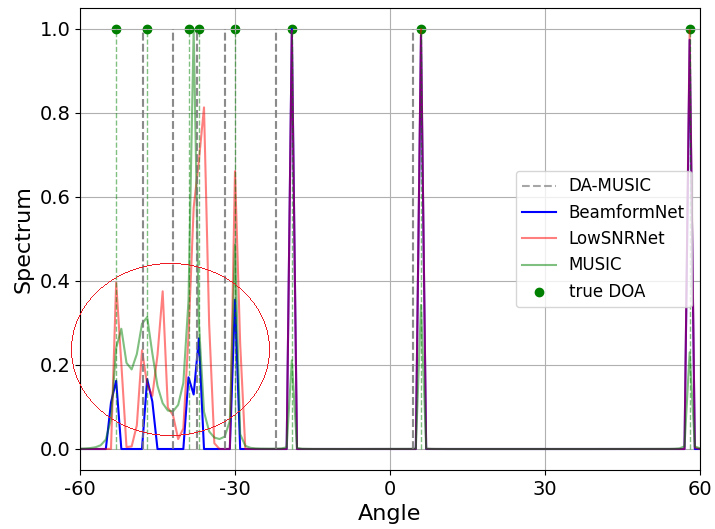}
        \caption{$K=8$}
        \label{fig:sub1}
    \end{subfigure}
    \hfill
    
    \caption{Spectrograms of various methods under different numbers of signal $K$}
    \label{fig:Intuitive_d}
\end{figure*}

\begin{figure*}[t]
    \centering
    \begin{subfigure}{0.45\linewidth}
        \centering
        \includegraphics[width=\linewidth]{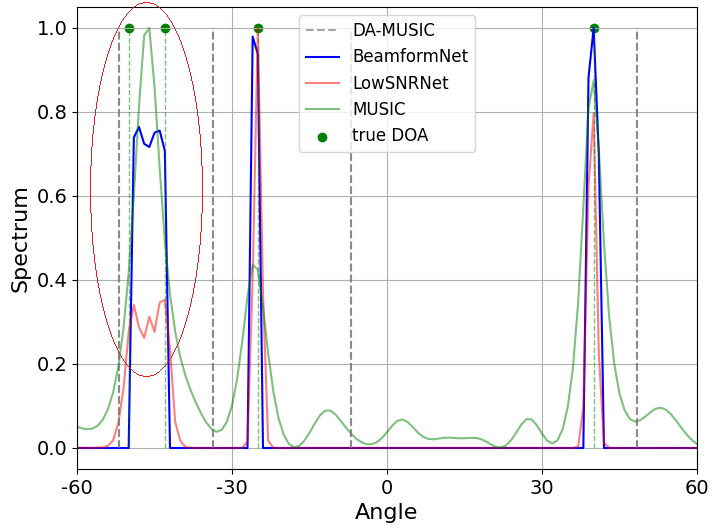}
        \caption{$SNR =-10dB$}
        \label{fig:sub1}
    \end{subfigure}
    \hfill
    \begin{subfigure}{0.45\linewidth}
        \centering
        \includegraphics[width=\linewidth]{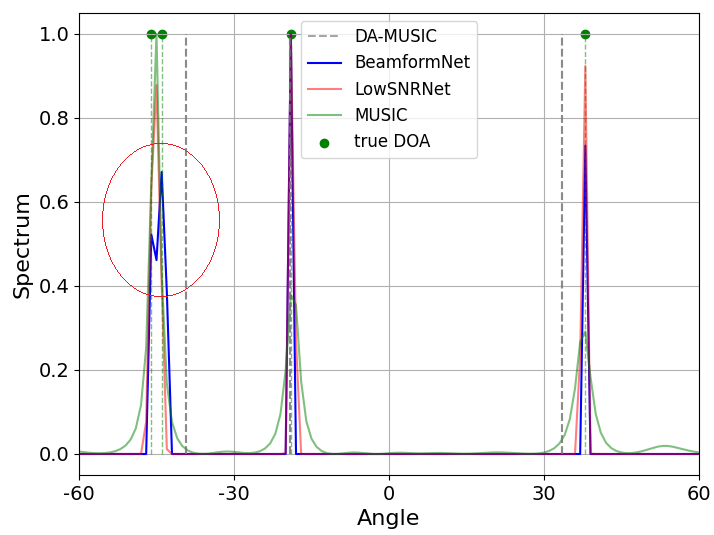}
        \caption{$SNR =-5dB$}
        \label{fig:sub1}
    \end{subfigure}
    \vfill
    \begin{subfigure}{0.45\linewidth}
        \centering
        \includegraphics[width=\linewidth]{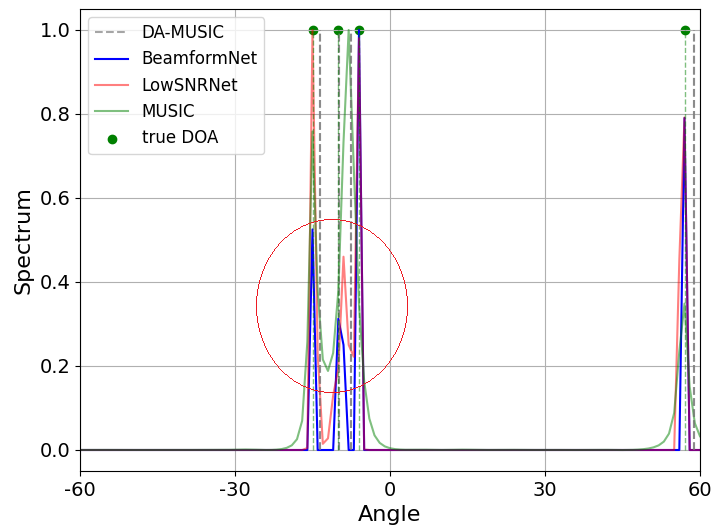}
        \caption{$SNR =0dB$}
        \label{fig:sub1}
    \end{subfigure}
    \hfill
    \begin{subfigure}{0.45\linewidth}
        \centering
        \includegraphics[width=\linewidth]{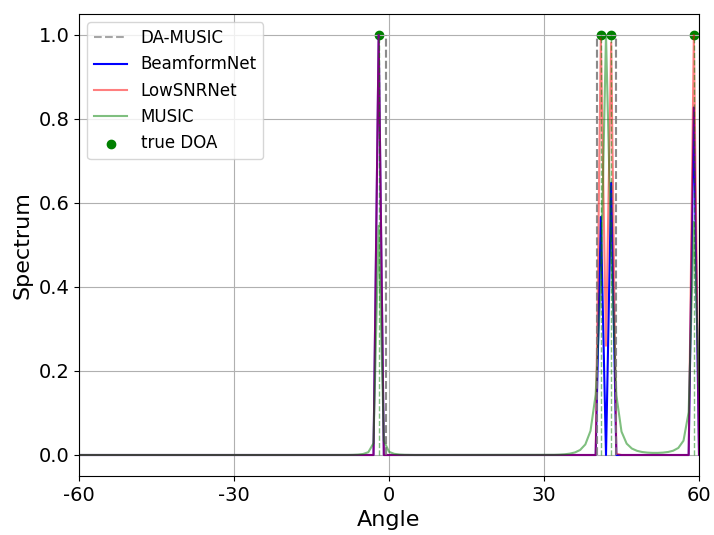}
        \caption{$SNR =5dB$}
        \label{fig:sub1}
    \end{subfigure}
    \hfill
    
    \caption{Spectrograms of various methods under different SNR}
    \label{fig:Intuitive_snr}
\end{figure*}

\begin{figure*}[t]
    \centering
    \begin{subfigure}{0.24\linewidth}
        \centering
        \includegraphics[width=\linewidth]{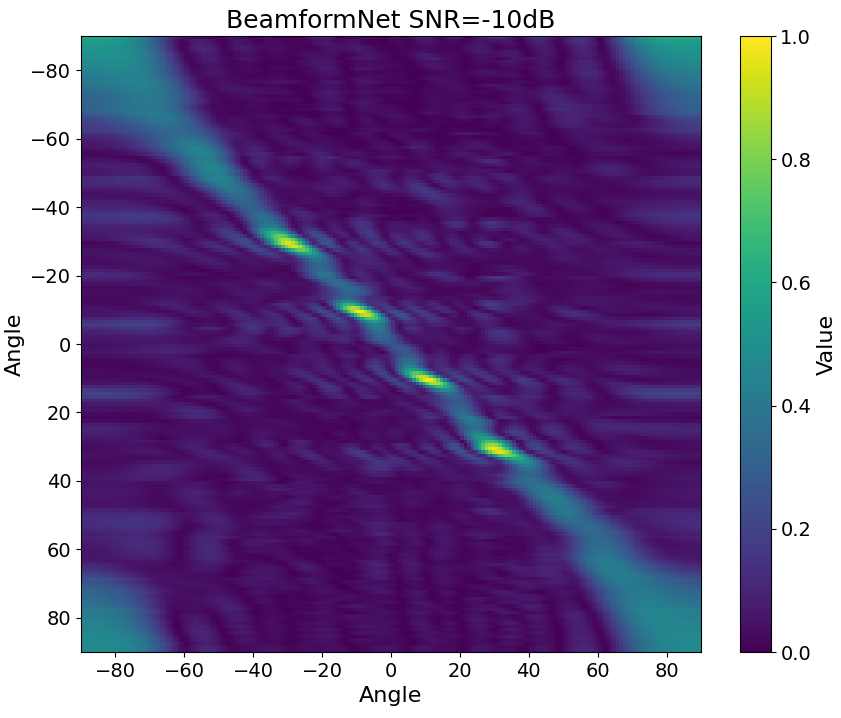}
        \caption{BeamformNet, SNR=-10dB}
        \label{fig:B_-10db_BeamformNet}
    \end{subfigure}
    \hfill
    \begin{subfigure}{0.24\linewidth}
        \centering
        \includegraphics[width=\linewidth]{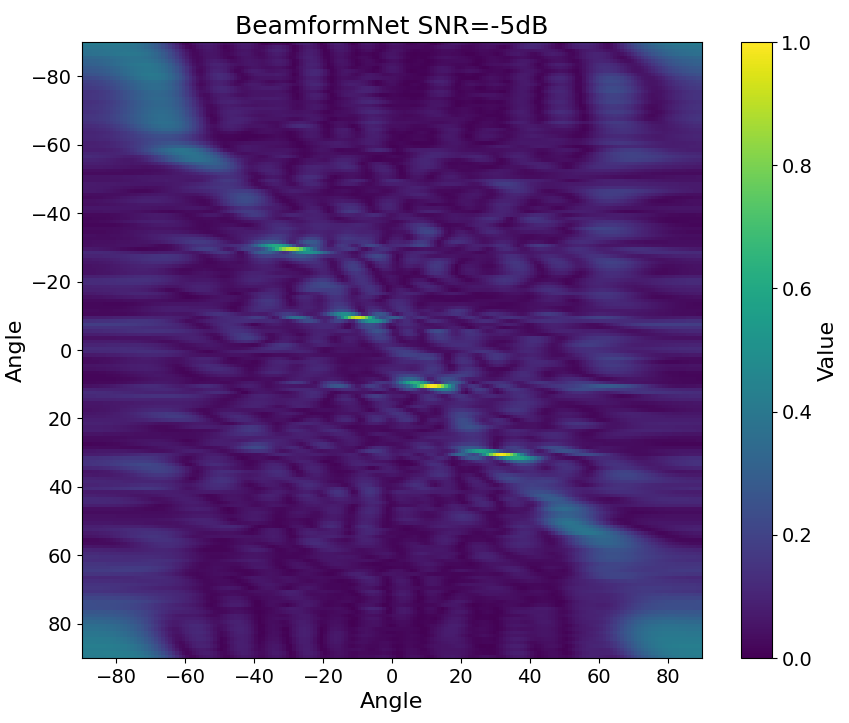}
        \caption{BeamformNet, SNR=-5dB}
        \label{fig:B_-5db_BeamformNet}
    \end{subfigure}
    \hfill
    \begin{subfigure}{0.24\linewidth}
        \centering
        \includegraphics[width=\linewidth]{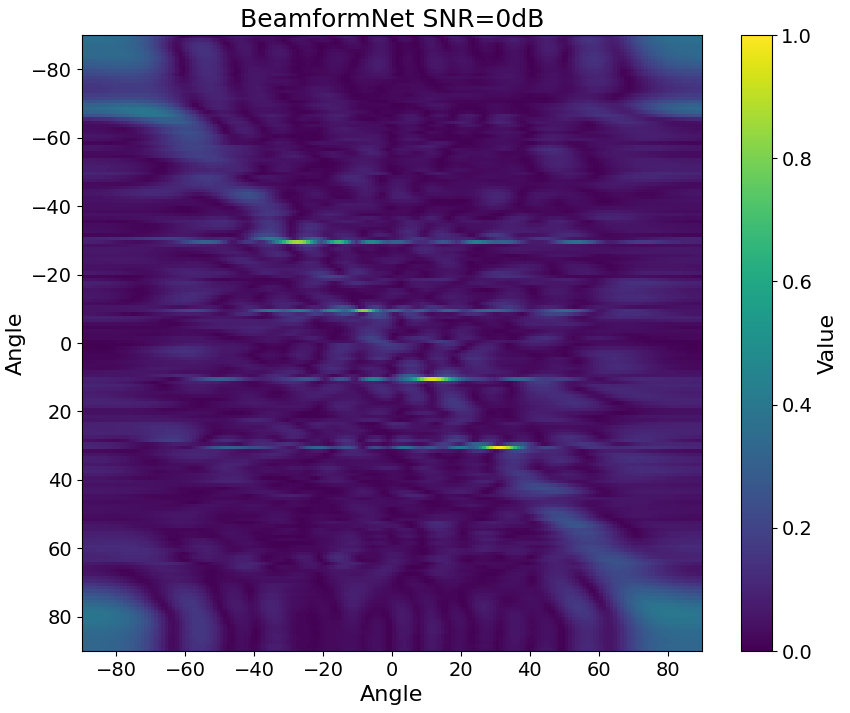}
        \caption{BeamformNet, SNR=0dB}
        \label{fig:B_0db_BeamformNet}
    \end{subfigure}
    \hfill
    \begin{subfigure}{0.24\linewidth}
        \centering
        \includegraphics[width=\linewidth]{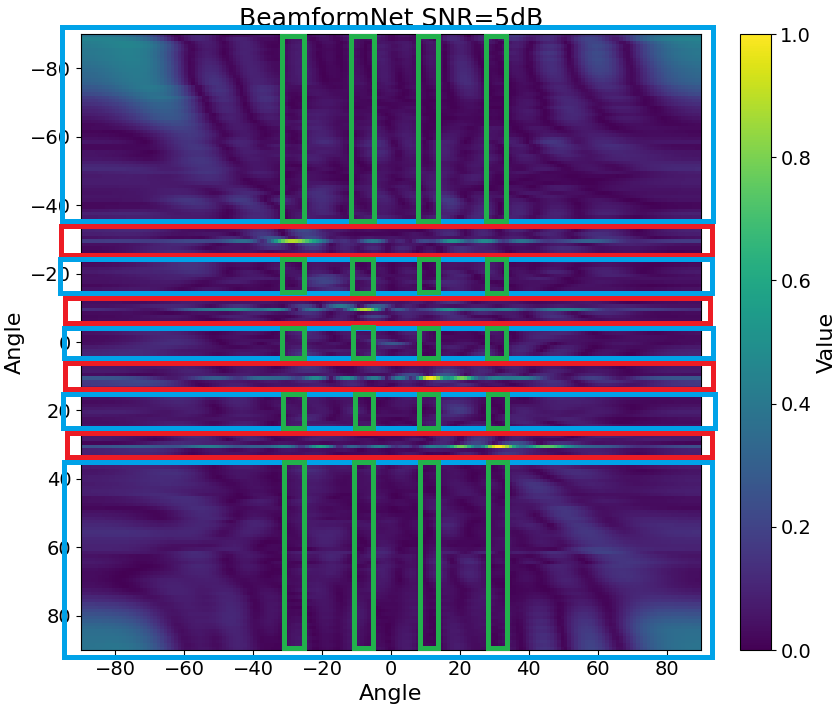}
        \caption{BeamformNet, $SNR=5dB$}
        \label{fig:B_5db_BeamformNet}
    \end{subfigure}
    \vfill
    \begin{subfigure}{0.24\linewidth}
        \centering
        \includegraphics[width=\linewidth]{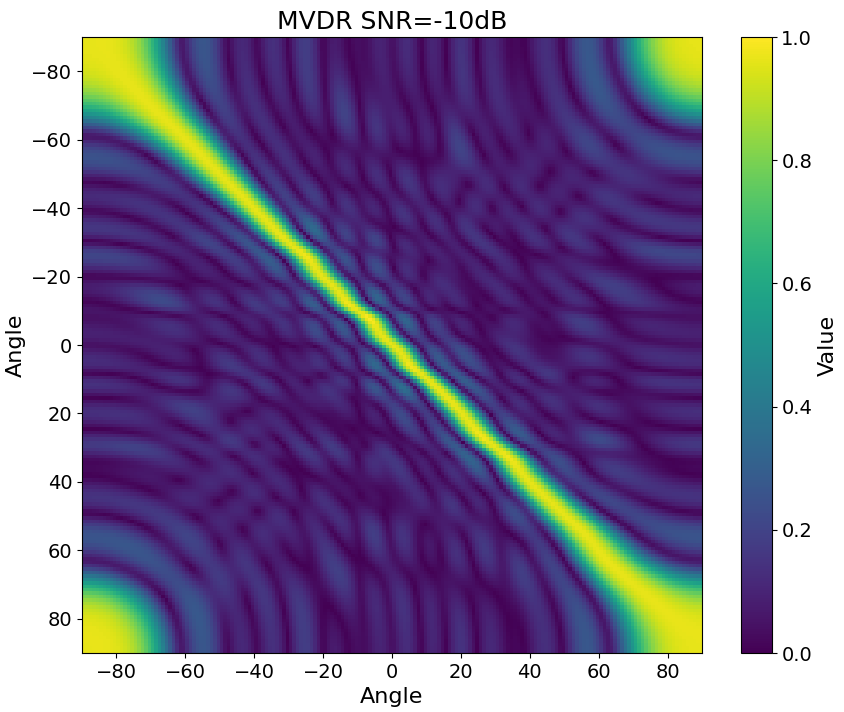}
        \caption{MVDR, SNR=-10dB}
        \label{fig:B_-10db_MVDR}
    \end{subfigure}
    \hfill
    \begin{subfigure}{0.24\linewidth}
        \centering
        \includegraphics[width=\linewidth]{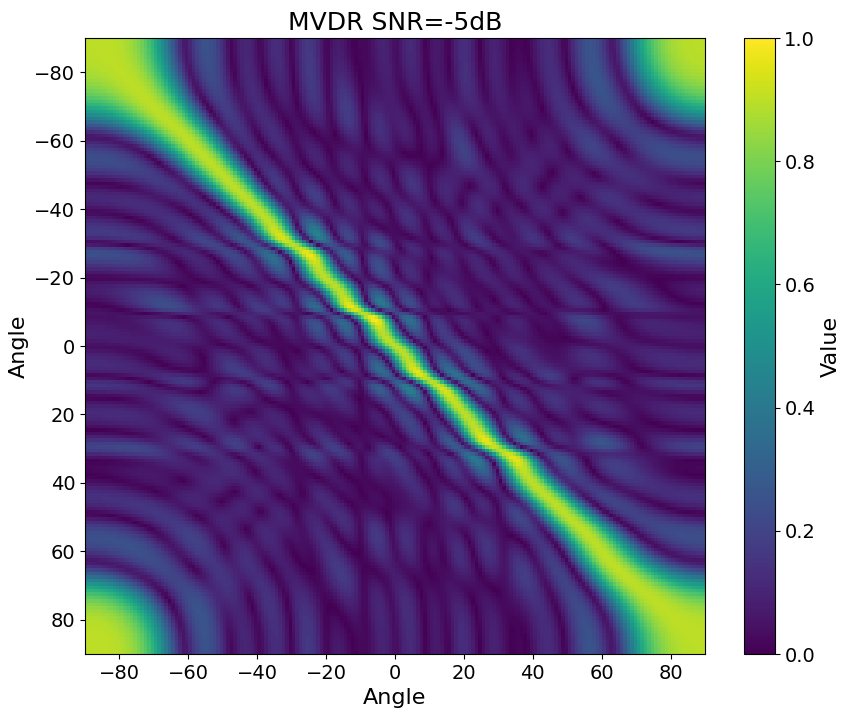}
        \caption{MVDR, SNR=-5dB}
        \label{fig:B_-5db_MVDR}
    \end{subfigure}
    \hfill
    \begin{subfigure}{0.24\linewidth}
        \centering
        \includegraphics[width=\linewidth]{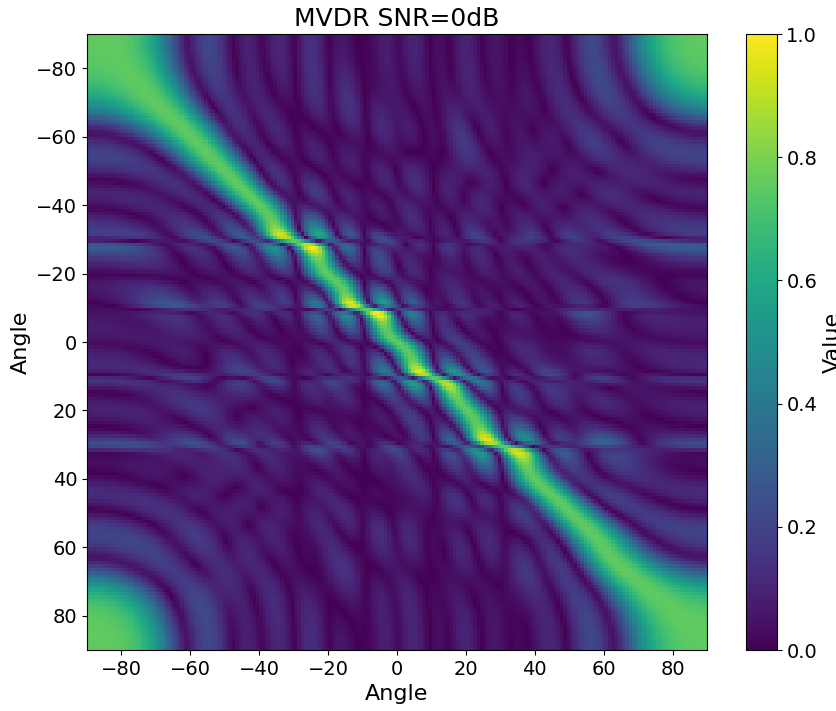}
        \caption{MVDR, SNR=0dB}
        \label{fig:B_0db_MVDR}
    \end{subfigure}
    \hfill
    \begin{subfigure}{0.24\linewidth}
        \centering
        \includegraphics[width=\linewidth]{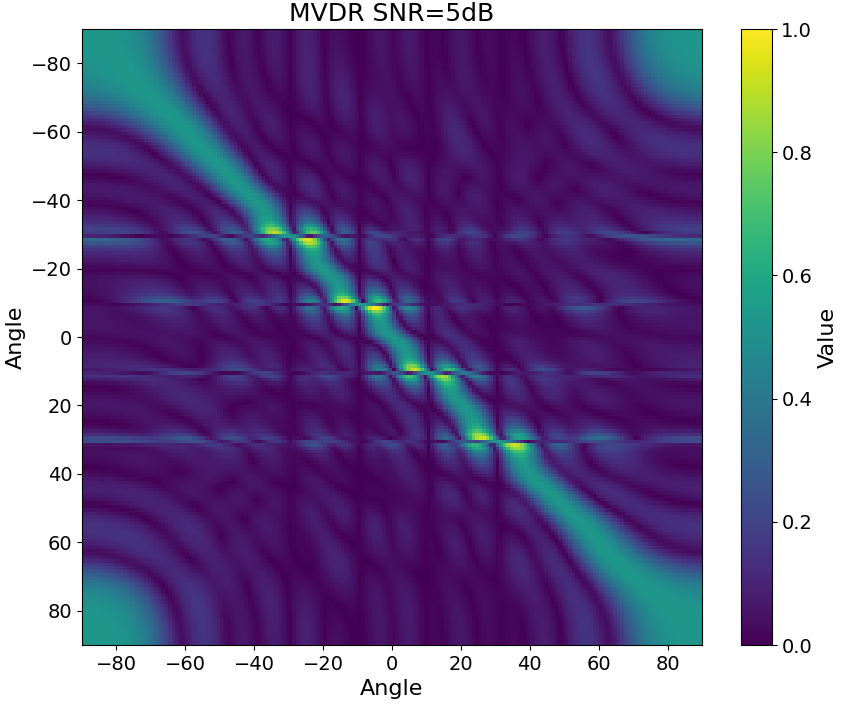}
        \caption{MVDR, SNR=5dB}
        \label{fig:B_5db_MVDR}
    \end{subfigure}
    \vfill
        \begin{subfigure}{0.24\linewidth}
        \centering
        \includegraphics[width=\linewidth]{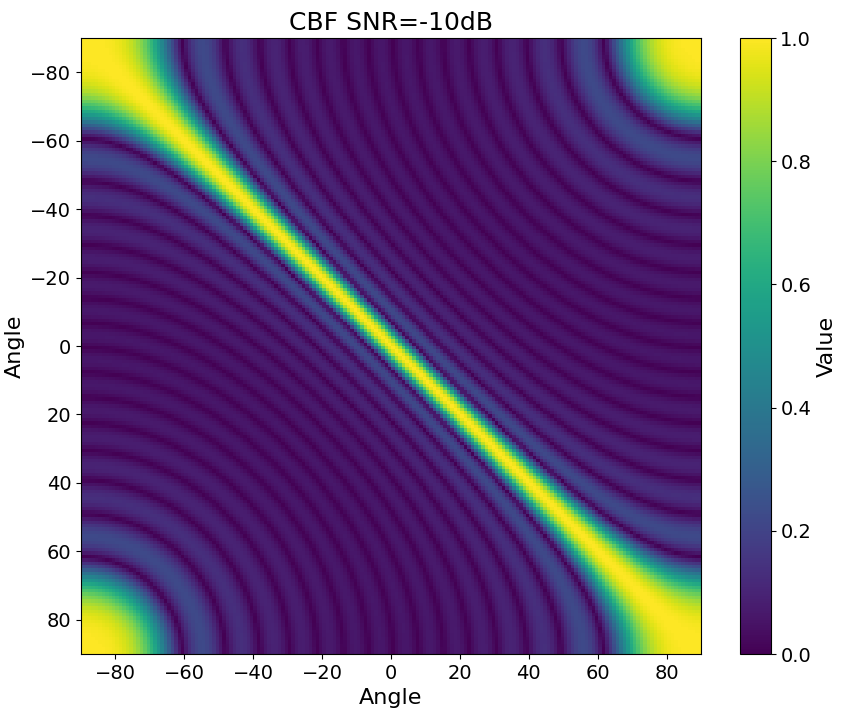}
        \caption{CBF, SNR=-10dB}
        \label{fig:B_-10db_CBF}
    \end{subfigure}
    \hfill
    \begin{subfigure}{0.24\linewidth}
        \centering
        \includegraphics[width=\linewidth]{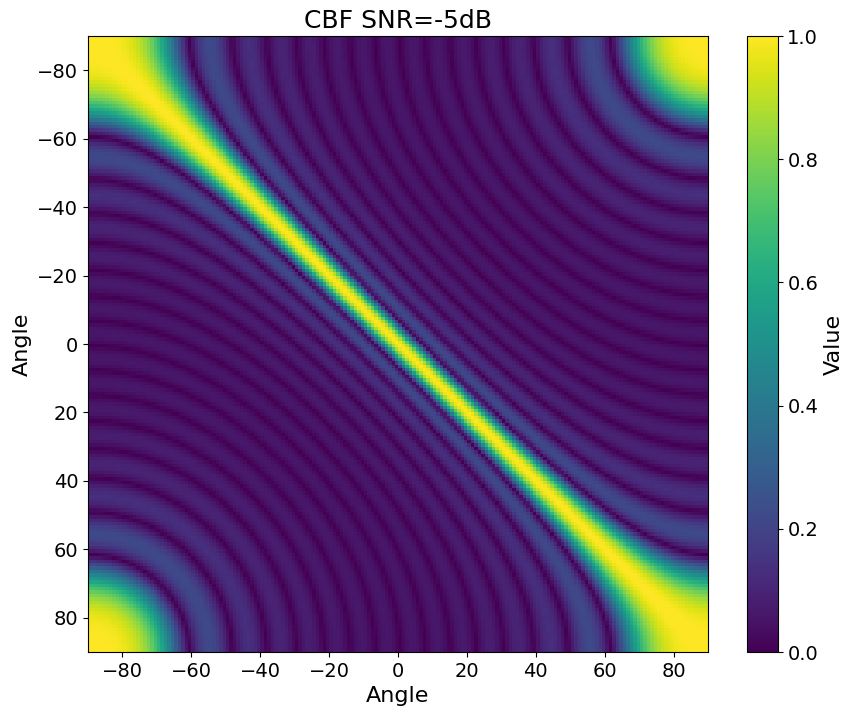}
        \caption{CBF, SNR=-5dB}
        \label{fig:B_-5db_CBF}
    \end{subfigure}
    \hfill
    \begin{subfigure}{0.24\linewidth}
        \centering
        \includegraphics[width=\linewidth]{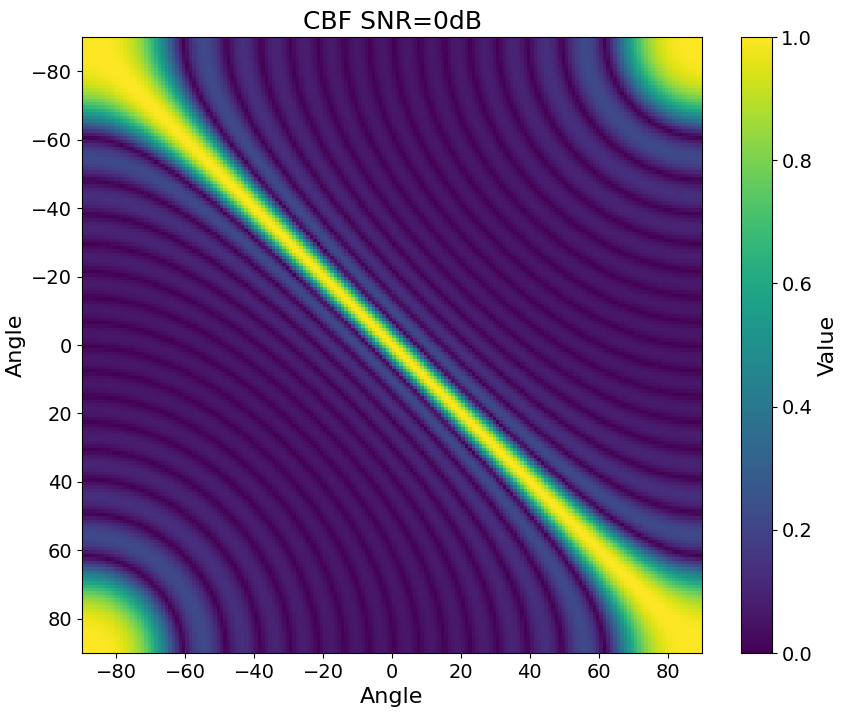}
        \caption{CBF, SNR=0dB}
        \label{fig:B_0db_CBF}
    \end{subfigure}
    \hfill
    \begin{subfigure}{0.24\linewidth}
        \centering
        \includegraphics[width=\linewidth]{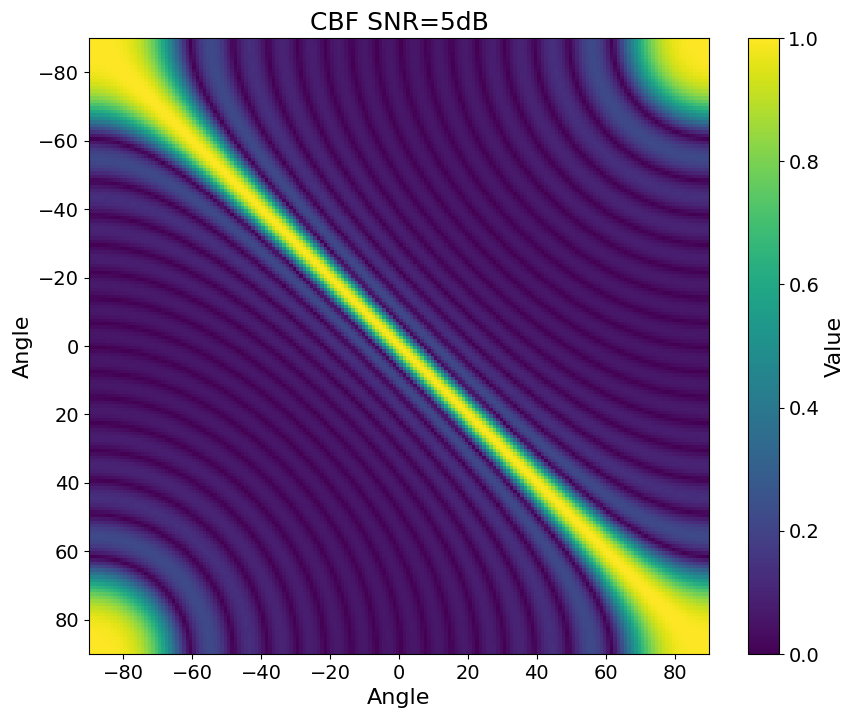}
        \caption{CBF, SNR=5dB}
        \label{fig:B_5db_CBF}
    \end{subfigure}
    \caption{This figure illustrates the spatial weighting matrix $\bm{W}$, which is obtained by multiplying the spatial filter $\bm{B}$ with a fixed array manifold matrix $\bm{A}$, where signals originate from directions $\{-30^\circ, -10^\circ, 10^\circ, 30^\circ\}$.}
    \label{fig:Intuitive_B}
\end{figure*}

\begin{table}[t]
\caption{Noise suppression performance for different methods measuring with SNR [dB].}
\label{tb:noise_suppression}
\begin{center}
\begin{small}
\begin{sc}
\begin{tabular}{@{}llll@{}}
\toprule
Method      & Original SNR & SNR after $B$ applied \\ \midrule
BeamformNet &  -15  &   \textbf{13.3018}       \\
MVDR &  -15  &   -2.6548  \\   
CBF &  -15  &   -2.9635      \\\bottomrule
\end{tabular}
\end{sc}
\end{small}
\end{center}
\end{table}

\subsection{Real-World Acoustic Source Localization}
Simulated signal data often differs significantly from real-world recordings. Real acoustic signals pose greater challenges due to multipath propagation, non-omnidirectional microphone responses, near-field broadband signals, and diverse environmental noise, among other factors. To evaluate BeamformNet on real-world data, we employed the RealMAN dataset~\cite{yang2024realman}, which contains speech recordings captured in realistic environments with ground-truth source positions. The RealMAN dataset comprises 83.7 hours of speech recordings (48.3 hours in static scenes and 35.4 hours in moving scenarios) collected across 32 distinct acoustic environments-including indoor, outdoor, semi-outdoor, and traffic settings-and 144.5 hours of background noise recorded in 31 different scenes. In the validation and test subsets, speech and noise signals have been pre-mixed, with average SNRs of 0.0 dB and -0.8 dB, respectively.

Due to the dataset including abundant samples, we only used the Moving subset as our real-world source localization dataset. From the RealMAN microphone array, we selected 10 microphones aligned along the $x$-axis (microphone indices: 25, 21, 13, 5, 0, 1, 9, 17, 26, 27) to form a uniform linear array. In our experiments, the audio signals were downsampled from 48 kHz to 16 kHz, and each data sample was constructed using a 30-ms time frame ($T = 480$). Ground-truth azimuth labels for each frame were obtained via linear interpolation. Additionally, voice activity detection (Google WebRTC) was applied to discard segments without active speech. Table~\ref{tb:realman} reports the source localization performance of various methods on this RealMAN subset. Notably, even under challenging near-field broadband conditions with real speech signals, BeamformNet consistently achieves the best performance among all evaluated approaches.

\subsection{The Impact of the Hyperparameter $E$}
\label{sec:E}

The embedding dimension of a model affects both its performance and size. Therefore, we investigate the impact of the hyperparameter $E$ on BeamformNet. Experiments are conducted under the following setup: $K=4, SNR=20 dB, T=200, M=16$, and incoherent signal sources. As shown in Figure 1, when the hyperparameter $E$ reaches 32, BeamformNet has already outperformed other methods and achieves optimal performance. And as the embedding dimension increases, the performance of BeamformNet continues to improve. We believe that a higher embedding dimension provides greater information capacity and representational flexibility, enabling BeamformNet to better learn complex spatio-temporal features and relationships from the input data. Furthermore, in Table 1, we list the number of parameters of BeamformNet as the hyperparameter $E$ varies, which offers flexible options for deploying BeamformNet in practical applications. When device capacity is the primary concern, a smaller embedding dimension can be chosen, yielding suboptimal yet acceptable performance. Conversely, when performance is the top priority, a model with a larger embedding dimension can be selected to achieve better results.

\subsection{Intuitive qualitative results}
\label{sec:Intuitive qualitative results}
To intuitively demonstrate the performance comparison between BeamformNet and other methods under the same conditions, we present spectrograms of various methods with different numbers of signals in Figure~\ref{fig:Intuitive_d} with $SNR=20, T=200, M=16$ and incoherent signal sources. Figure~\ref{fig:Intuitive_snr} shows spectrograms of various methods under different SNR ratios with $K=4, T=200, M=16$ and incoherent signal sources. Figures~\ref{fig:Intuitive_d} and~\ref{fig:Intuitive_snr} have shown that BeamformNet achieves higher resolution accuracy and exhibits better performance in DoA estimation.

Additionally, we visualized the spatial filter $\bm{B}$ obtained by BeamformNet through the array manifold matrix $\bm{A}$. As shown in Figure~\ref{fig:Intuitive_B}, the spatial filter $\bm{B}$ multiplied by the array manifold matrix $\bm{A}$ yields weights $\bm{W}$ that weigh different spatial grids. We visualized the weighting matrix $\bm{W}$ for different methods when signals originate from directions $\{-30^\circ, -10^\circ, 10^\circ, 30^\circ\}$. For BeamformNet and MVDR, the weighting matrix $\bm{W}$ is averaged over 1000 trials with random data ($T=200$). A desirable weighting matrix $\bm{W}$ should assign high weights only to grids corresponding to the actual signal directions, while assigning low or zero weights elsewhere. It can be observed that BeamformNet’s weighting matrix $\bm{W}$ assigns high weights to grids where signals are present (row vectors, red boxes), thereby avoiding missed signals and providing sharper and more refined responses. For grids without signals (row vectors, blue boxes), the weights in grids containing signals (green boxes) remain low, which prevents introducing false signals into grids where they do not exist and reduces false alarms. This also explains why BeamformNet can separate spatial signals in Formula~\eqref{eq:BAS=S}. In contrast, traditional model-driven methods such as MVDR and CBF yield coarser weighting matrices $\bm{W}$. Since CBF is non-adaptive, its weighting matrix $\bm{W}$ is unaffected by SNR and shows a constant response, equivalent to scanning different spatial directions. MVDR performs well at high SNRs, but begins to degrade when SNR falls below 0 dB, exhibiting a response similar to that of CBF. BeamformNet, however, demonstrates more robust responses to spatial signals at low SNRs.

In Formula~\eqref{eq:BN=0}, another role of the spatial filter $\bm{B}$ is noise suppression. We computed the signal-to-noise ratio before and after applying the spatial filter $\bm{B}$ to the noise matrix $\bm{N}$. As shown in Table~\ref{tb:noise_suppression}, it can be observed that BeamformNet exhibits superior noise suppression performance compared to traditional model-driven beamforming algorithms. Even at low SNR, it can enhance the SNR from -15 dB to 13.3 dB by effectively suppressing noise.

\section{Conclusion}
%In this paper, we propose a novel method, BeamformNet, for DoA estimation. The BeamformNet architecture is inspired by model-driven beamforming algorithms, and the learning framework of BeamformNet is theoretically grounded to approximately achieve optimal DoA estimation by approximating the optimal spatial filter. Extensive experimental results demonstrate that BeamformNet shows robustness and outperforms existing methods across a wide range of scenarios—both in simulated environments and in challenging real-world acoustic conditions involving near-field broadband speech signals. These findings validate the effectiveness of approximating the optimal solution through deep neural network-based learning. Future work will focus on further improving the model’s performance under extreme conditions, such as single-snapshot and very low-SNR scenarios.
This paper introduces BeamformNet, a novel deep learning-based beamforming method for DoA estimation. Drawing inspiration from traditional model-driven beamforming techniques, BeamformNet's architecture and learning paradigm are theoretically founded on the approximation of an optimal spatial filter, enabling implicit signal focusing and noise suppression to achieve near-optimal DoA accuracy.
Comprehensive experiments on simulated datasets and real-world near-field broadband speech localization benchmarks validate BeamformNet's superior robustness and performance, surpassing state-of-the-art methods across diverse scenarios, including varying numbers of sources, snapshots, SNRs, angular separations, array elements, and mismatches—even under coherent sources and limited snapshots.
These results underscore the efficacy of leveraging deep neural networks to approximate optimal solutions in array signal processing. Future research will aim to enhance the model's capabilities in extreme conditions, such as single-snapshot and ultra-low-SNR environments.

%introduction里记得加匿名GitHub

\bibliographystyle{IEEEtran}
\bibliography{refs}

\begin{IEEEbiography}[{\includegraphics[width=1in,height=1.25in,clip,keepaspectratio]{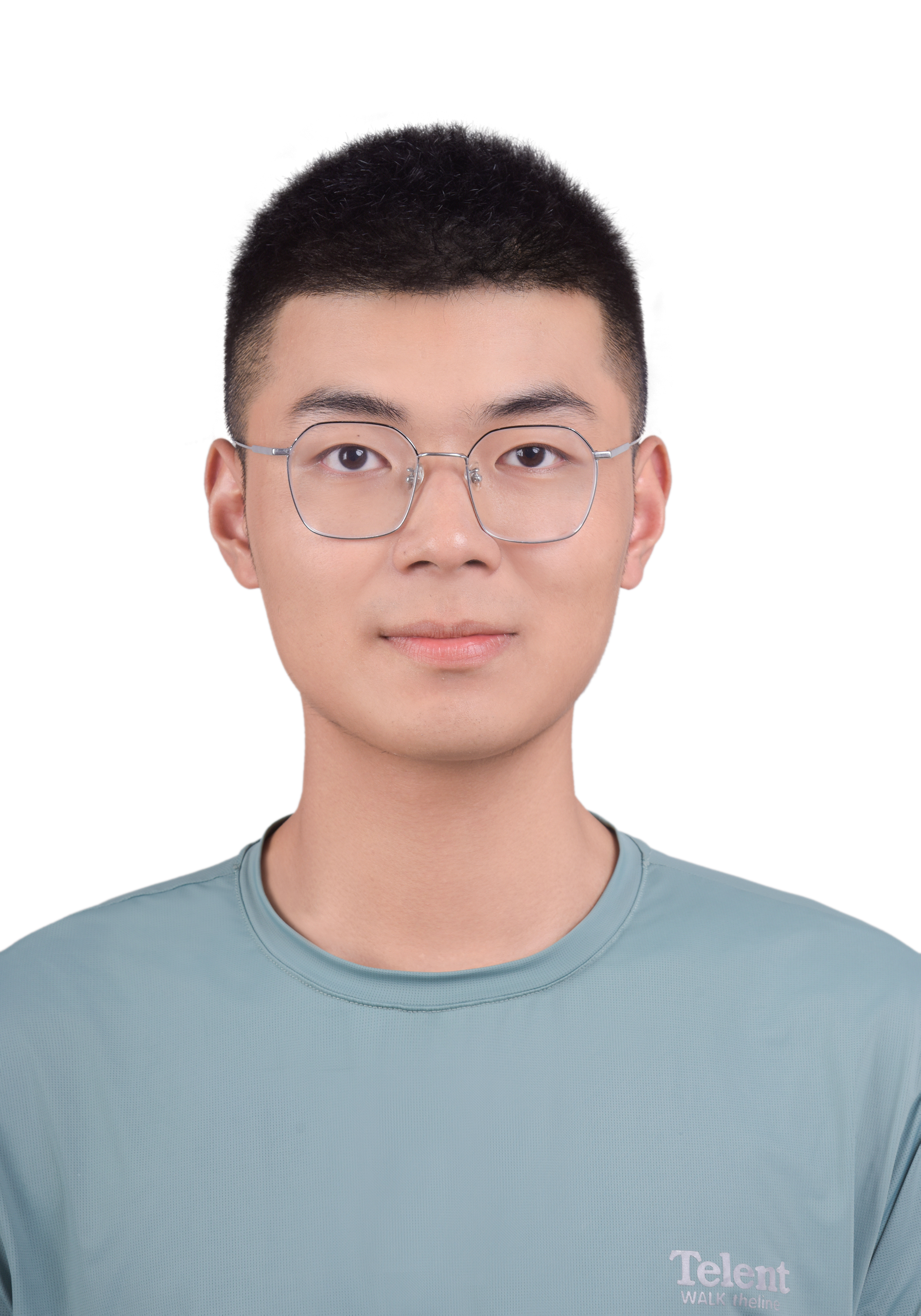}}]{Xuyao Deng} received the bachelor's degree from Shandong University, Shandong, China, in 2023. He is currently a doctoral student at the School of Computer Science, National University of Defense Technology, Changsha, China. His research interests include signal processing, machine learning, and intelligent software systems.
\end{IEEEbiography}

\begin{IEEEbiography}[{\includegraphics[width=1in,height=1.25in,clip,keepaspectratio]{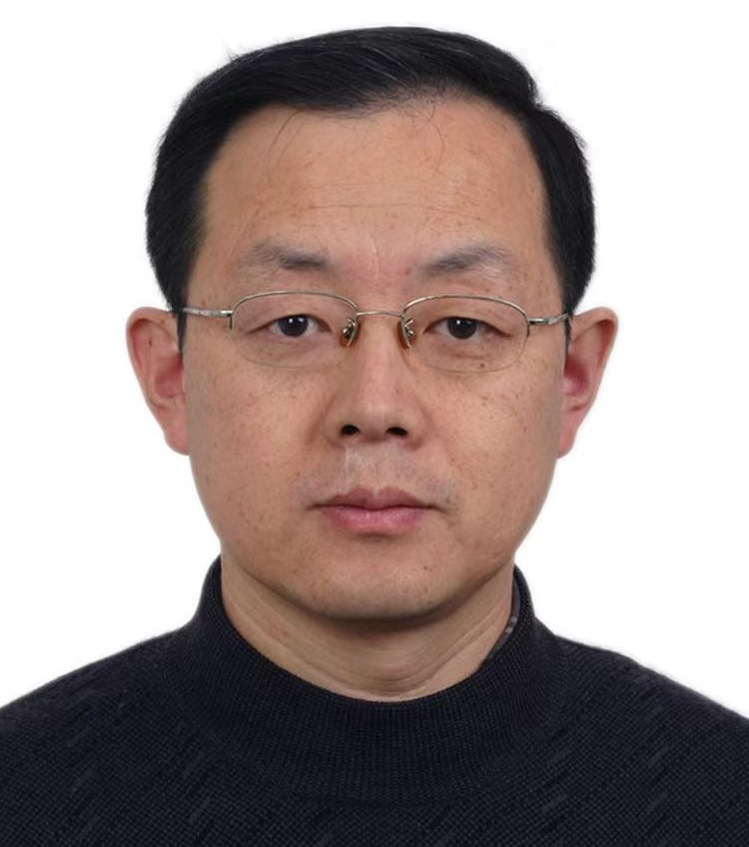}}]{Yong Dou}
Professor and Ph.D. supervisor with the National Key Laboratory of Parallel and Distributed Computing, National University of Defense Technology. His research interests cover high performance computing, intelligent computing, machine learning, and deep learning.
\end{IEEEbiography}

\begin{IEEEbiography}[{\includegraphics[width=1in,height=1.25in,clip,keepaspectratio]{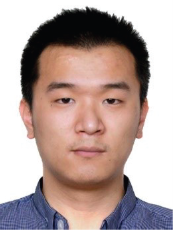}}]{Kele Xu}
(Senior Member, IEEE) received the doctorate degree from Paris VI University, Paris, France, in 2017. He is currently an Associate Professor with the School of Computer Science, National University of Defense Technology, Changsha, China. His research interests include audio signal processing, machine learning, and intelligent software systems. He is the associate editor for IEEE Transactions on Circuits and Systems for Video Technology and Guest editor for Science Partner Journal Cyborg and Bionic Systems. He has (co-)authored more than 100 publications in peer reviewed journals and conference proceedings, including IEEE TASLP, TMI, ICML, CVPR, NeurIPS, ICLR, AAAI, IJCAI, SIGIR, ASE, ACM MM, ICASSP.
\end{IEEEbiography}

\end{document}